\documentclass[pre,showpacs,showkeys,preprintnumbers,amsmath,amssymb,superscriptaddress,twocolumn]{revtex4-1}
\usepackage[english]{babel}
\usepackage[dvipsnames]{xcolor}
\usepackage{makeidx} % Produzir indice remissivo
\usepackage{graphicx} % Incluir graficos no arquivo.
\usepackage{dcolumn} %Permitir alinhamento decimal em tabelas. Precisa do Array Package.
\usepackage{array} % Adicionar novas particularidades para as tabelas e vetores.
\usepackage{amssymb} %Permite o uso de sofisticados simbolos matematicos
\usepackage{amsmath}
\usepackage{textcomp}
\usepackage{multirow}
\usepackage{subfigure}
\usepackage{eucal}
\usepackage{mathrsfs}
\usepackage{dsfont}
\usepackage{physics}
\usepackage[utf8]{inputenc}
\usepackage{amsfonts} %????
\usepackage{bm} %????
\usepackage{float}

\begin{document}

\title{Measurement-based quantum heat engine in a multilevel system}
 
\author{Maron F. Anka\email{maronanka@id.uff.br}}
\affiliation{Instituto de F\'{i}sica Universidade Federal Fluminense - Av. Gal. Milton Tavares de Souza s/n, 24210-346 Niter\'{o}i, Rio de Janeiro, Brazil.}
\author{Thiago R. de Oliveira} 
\affiliation{Instituto de F\'{i}sica Universidade Federal Fluminense - Av. Gal. Milton Tavares de Souza s/n, 24210-346 Niter\'{o}i, Rio de Janeiro, Brazil.}
\author{Daniel Jonathan} 
\affiliation{Instituto de F\'{i}sica Universidade Federal Fluminense - Av. Gal. Milton Tavares de Souza s/n, 24210-346 Niter\'{o}i, Rio de Janeiro, Brazil.}

\date{\today}
 
\begin{abstract}
	
We compare quantum Otto engines based on two different cycle models: a two-bath model, with a standard heat source and sink, and a measurement-based protocol, where the role of heat source is played by a quantum measurement. We furthermore study these cycles using two different `working substances': a single qutrit (spin-1 particle) or a pair of qubits (spin-1/2 particles) interacting via the XXZ Heisenberg interaction. Although both cycle models have the same efficiency when applied on a single-qubit working substance, we find that both can reach higher efficiencies using these more complex working substances, by exploiting the existence of `idle' levels, i.e., levels that do not shift while the spins are subjected to a variable magnetic field.  Furthermore, with an appropriate choice of measurement, the measurement-based protocol becomes more efficient than the two-bath model.

\textsl{}\end{abstract}

\maketitle

\section{INTRODUCTION}

The fast development of new technologies into the nano scale makes quantum effects no longer negligible. Thus there is a natural need to understand which, if any, quantum effects may enhance thermodynamic processes. This emerging subject, known as quantum thermodynamics \cite{binder2019thermodynamics}, has received attention from different fields, such as quantum information \cite{sagawa2012thermodynamics,yi2017role,goold2016role,parrondo2015thermodynamics,goold2016role,sagawa2013information,sagawa2019second,sagawa2012thermodynamics}, quantum optics \cite{youssef2009quantum,de2018experimental,klatzow2019experimental,passos2019optical,zanin2019experimental} and resource theory \cite{renes2014work,brandao2015second}. 

A main focus of this field has been the study of quantum heat engines (QHEs), i.e, heat engines that work with a small quantum system as the working substance. Recently, several such models have been analyzed, such as those based on spin systems \cite{zhang2008entangled,thomas2011coupled,altintas2015quantum,ccakmak2017special,ccakmak2017special,zhao2017entangled,he2012thermal,thomas2017implications,chand2017measurement,hewgill2018quantum,ivanchenko2015quantum,mehta2017quantum,ding2018measurement}, harmonic oscillators \cite{deffner2008nonequilibrium,horowitz2012quantum,wang2007performance,schnack1999thermodynamics,hilt2009system}, trapped ions \cite{bushev2006feedback,rossnagel2016single}, and others \cite{quan2007quantum,quan2009quantum,wang2009thermal,kosloff2014quantum,huang2018multilevel}. We should mention that already in 1959 a three-level maser was analyzed as a thermal engine \cite{scovil1959three}.

One of the main issues when treating small quantum systems thermodynamically is deciding how to classify energy transfer processes as heat or work. In the classical macroscopic scenario it is clear that heat is a process where energy is transferred by microscopic degrees of freedom in uncontrollable, random ways and associated with entropy production. Work, on the other hand, is the energy transfer through a macroscopic and controllable degree of freedom. Such a difference is fundamental, since thermodynamics' main practical concern is how to convert heat that is "freely" available in nature into work, and that is the purpose of a heat engine. 

There are already some definitions that are used in most analyses of QHE,
but with no clear picture of the role of entanglement or other quantum correlations in the operation of the engine
\cite{geusic1967quantum,quan2007quantum,quan2009quantum,wang2013quantum,kosloff2014quantum, klatzow2019experimental,dorfman2018efficiency,camati2019coherence,huang2013special,he2012thermal,zhang2008entangled,huang2012effects}.

Recently it was proposed that, since a quantum measurement is a random and irreversible process, any energy it transfers to or from the measured system may be considered as a form of heat (quantum heat)  Thus one can have a new kind of engine with one of the thermal baths replaced by a quantum measurement \cite{elouard2017extracting,strasberg2019operational,jordan2020quantum,monsel2020energetic,auffeves2021short}. In this scenario one can furthermore consider using either selective measurements and feedback (a Maxwell demon \cite{maxwell2001theory,szilard1964decrease,leff2014maxwell,kim2011quantum,koski2014experimental,elouard2017extracting}) or non-selective measurements and no feedback \cite{yi2017single,ding2018measurement,das2019measurement},
\footnote{We should not confuse these engines with Szilard-type engines. There the measurement, done by a Maxwell demon, is used to extract only information about the system, that can then be used to extract work. 
Even though there could be some energy exchange during the information extraction itself, this is not the main purpose, a point made by David Deutsch. There are many analyses of Szilard-type engines in the quantum scenario \cite{quan2006maxwell,kim2011quantum,zurek1986maxwell,PhysRevLett.118.260603,esposito2011second,hilt2011landauer,berut2012experimental}, some even considering entanglement and other quantum correlations \cite{wang2009thermal,zhang2008entangled,zhang2007four,scully2011quantum}.}. In the latter case, the post-measurement state is taken as the appropriate average over all possible measurement results. In other words, the measurement is treated as a quantum channel. In this article we will only consider this scenario, and we will use the expressions `quantum measurement' and `quantum channel' interchangeably. We should
mention that there are also proposals where the measurement acts as an external
source of work \cite{Buffoni19}

Yi et al.  \cite{yi2017single} studied the use of such a non-selective measurement in an Otto-type cycle. This cycle is composed of two quantum adiabatic strokes, where no heat transfer occurs, one conventional isochoric stroke, where the system equilibrates with an external cold reservoir without realizing work, and one measurement stroke, which plays the role of the hot thermal reservoir. They pointed out that measurements whose Kraus operators are all Hermitian (in particular, projective measurements) always increase the energy of the system in this cycle, and can thus play the role of the engine's heat source. They then showed that, for working substances whose energy gaps all vary by the same ratio in the adiabatic strokes, such as a harmonic oscillator or non-interacting qubits, the cycle's efficiency is independent of the specific measurement made, and has in fact the same value as when these systems interact with a conventional hot thermal bath. Thus at least in these simple examples, a measurement-based engine has no advantage over a conventional two-bath one. 

Das and Ghosh \cite{das2019measurement} later investigated whether this conclusion still holds in more complex scenarios. They studied  the same measurement-based protocol in the cases of two coupled qubits, and of one qubit coupled with a general spin $S$, interacting via an isotropic Heisenberg Hamiltonian. They found that the efficiency can indeed be greater in these cases, but it was not clear if this was an effect of the coupling, of the measurement, or both. They also did not compare their results with the corresponding efficiency values of an engine with the same coupled qubits but using two thermal baths.

Recently, some of us have shown \cite{Oliveira20} that efficiency gains in this corresponding two-qubit, two-bath model come not from entanglement or any quantum correlation, but just from the way the various system levels channel heat during the cycle. More specifically, an efficiency increase is possible when some levels are `idle', in the sense that they do not couple to the external work sink, i.e, they do not shift during the adiabatic strokes. Any heat absorbed from one bath by these levels cannot be converted into work, and must be deposited in the other bath. Channeling this heat flow from the cold to the hot bath allows more heat to flow in the opposite sense through the coupled levels, thus increasing efficiency.

Here we seek to better understand if and how a measurement-based engine without feedback can have greater performance than the corresponding two-thermal-bath model.  With this aim we first study and compare both kinds of  Otto cycle in the case of a toy model with just three energy levels. Since this qutrit system is not divisible into subsystems, there is no question of any efficiency gains being due to entanglement or other quantum correlations. 

As in Ref. \cite{Oliveira20}, we find that efficiency gains (relative to the uncoupled-qubits case) occur when one of the qutrit levels is `idle'. 
%We find that the origin of the efficiency increase is not related to entanglement or any quantum correlation, but to the way the various levels channel heat, as also recently shown by some of us in the case of the two-bath model itself \cite{Oliveira20}. More specifically, an efficiency increase is possible due to the fact that one of the three levels does not couple to the external work sink, while the other two are. Thus heat absorbed by this level cannot be converted into work. Channeling this heat flow from the cold to the hot bath allows more heat to flow in the opposite sense through the coupled levels, thus increasing efficiency.
Moreover, we find that the measurement-based protocol allows a fine-tuning of the reversed heatflow mechanism, which can result in even larger efficiencies than in the corresponding two-bath model. Finally we explore the same possibility in the case of two qubits interacting via an anisotropic Heisenberg Hamiltonian.

The structure of the paper is as follows: In section~II we briefly review quantum Otto engines composed of two quantum adiabatic strokes and two isochoric interactions with thermal reservoirs. In section~III we review how the hot  reservoir may be replaced by a non-selective quantum measurement, and we show that in fact any unital measurement can be used in this way. In section~IV we apply both protocols to a qutrit system and find the conditions for achieving an efficiency higher than the qubit limit  $\eta_0 = 1 - 1/r$, where $r = \lambda_{\max}/\lambda_{\min}$ is the ratio between the higher and lower values of the adiabatic parameter $\lambda$. Next, in section~V, we use two interacting qubits as our working substance and study the role of the interaction in the efficiency. Finally, we conclude by explaining the mechanisms behind the increase in efficiency.

\section{QUANTUM OTTO ENGINE}

In this section we review the Otto heat engine model in the quantum regime. Let us first briefly introduce the concepts of work and heat in the quantum setting. We first define the internal energy of the system as the energy expectation value
\begin{equation}
U = \langle \mathcal{H} \rangle = Tr[\rho \mathcal{H}].
\end{equation}
Then, in any infinitesimal process we can state the first law of thermodynamics for quantum  systems as \cite{vinjanampathy2016quantum,deffner2019quantum,kosloff2013quantum,alicki2004thermodynamics,seifert2016first}:
\begin{equation}
\begin{split}
dU & = Tr[d\rho \mathcal{H}] + Tr[\rho d\mathcal{H}] \\
& = \delta Q  + \delta W ,
\end{split}
\end{equation}
where $dU$ is the change in the system's average energy, and we have defined 
work, $W $, as the change in the average energy due to a change in external and controllable parameters of $\mathcal{H}$. Heat, $Q $, on the other hand, is defined as the change in the average energy when
all controllable parameters are fixed; these changes usually come
from interactions with the environment. It is worth mentioning that these interpretations are valid only for weak system-bath  coupling  \cite{kosloff2013quantum,alicki2004thermodynamics,seifert2016first}. 

%%% Nota DJ: tirei os colchetes de 'valor esperado' de Q e W

A standard thermal engine consists of a `working substance' (WS), a system
that undergoes a cycle during which it interacts with two thermal baths at different temperatures. Through this cyclic process, some of the heat flowing between the baths is converted into work. Let us consider that our WS is a generic quantum system governed by the Hamiltonian 
\begin{align} 
\mathcal{H}(\lambda)=\sum_n E_n(\lambda) |E_n(\lambda)\rangle \langle E_n(\lambda)|
\end{align}
with $\lambda$ some tunable external parameter.
We will also assume that after sufficient time in contact with a thermal
bath at inverse temperature $\beta = 1/(k_{B}T)$, the system reaches the corresponding Gibbs thermal equilibrium state, $\rho(\lambda) = e^{-\beta \mathcal{H}(\lambda)}/Z(\lambda)$, with $Z(\lambda) = Tr[e^{-\beta \mathcal{H}(\lambda)}]$ the partition function.

The Otto cycle consists of four processes (strokes): two adiabatic ones, where there is no heat exchange and two ``isochoric" \cite{Fermi56} ones, where there is no work exchange.

First stroke: The first stroke is an isochoric process in which the working substance thermalizes with a cold heat bath at inverse temperature $\beta_c = 1/(k_BT_c)$. No work is done in this step, since
$\lambda$ is fixed at $\lambda_i$, and only heat is released by the system into the bath. Following the definition given above, this heat exchange is 
\begin{equation} \label{eq:Qc}
 Q_c  = \sum_n E_n(\lambda_i)(p_n^c - p_n^h).
\end{equation}
where $p_n^{c(h)} = \exp[-\beta_{c(h)} E_n(\lambda_{i(f)})] / Z(\lambda_{i(f)})$ are occupation probabilities of the $n^{th}$ energy level of the system in thermal equilibrium with the cold (hot) bath, described by the density matrix $\rho(\lambda_{i(f)})$.

Second stroke: In this step, we detach the working substance from the cold heat bath and let the external parameter adiabatically change from its initial value to a final one, $\lambda_i \rightarrow \lambda_f$. Thus, the only contribution to energy change is in the form of work; there is no heat. In this paper, we assume for simplicity that this is a true quantum adiabatic evolution, where there is no change in the energy occupation probabilities, whereas the energy eigenvalues and eigenstates evolve smoothly from those of $\mathcal{H}(\lambda_{i})$ to those of $\mathcal{H}(\lambda_{f})$ \footnote{This is not always the same as a thermodynamic adiabatic process. In the quantum case 
% the process may  be irreversible, due to the fact that the
the system does not necessarily remain in an equilibrium thermal state of its respective $\mathcal{H}$ \cite{quan2006multilevel}.}. 
% the process may  be irreversible, due to the fact that the
This requires the timescale for the change in $\lambda$ to be at least as large as the inverse of the smallest relevant energy gap. In particular, no level crossings can occur.  It is important to note that, although the probabilities $p_{n}^{c}$ at the end of this step are therefore those of the thermal state $\rho(\lambda_{i})$, the corresponding state will generally \emph{not} be thermal with respect to the final Hamiltonian $\mathcal{H}(\lambda_{f})$. It is still however a `passive' state,  i.e., an energy-diagonal state where the probabilities $p_{n}$ are monotonically nonincreasing with energy $E_{n}$, and thus it contains no ergotropy \cite{allahverdyan2004ergotropy}.

%This process is analogous to an adiabatic compression of a gas. 

Third stroke: This is another isochoric process in which the working substance is put in thermal contact with a hot heat bath at inverse temperature $\beta_h=~1/(k_BT_h)<\beta_{c}$. After the system thermalizes with the bath, it absorbs the energy 

\begin{equation} \label{eq:Qh}
Q_h = \sum_n E_n(\lambda_f)(p_n^h - p_n^c).
\end{equation}
Again, no work is done during this process.

Fourth stroke: This process is similar to the second stroke. Here, the external parameter in changed back to the initial value, $\lambda_f \rightarrow \lambda_i$, and the occupation probabilities remain fixed at $p_n^h$. Only work is performed and no heat is exchanged.

Due to energy conservation in a cyclic process, the total work done by the system is equal to the negative sum of the total heat transferred during steps 1 and 3:

\begin{equation}
\begin{split}
W &= - (Q_h + Q_c) \\
&= -\sum_n \Delta E_n \Delta p_n
\end{split}
\end{equation}
where $\Delta E_n = E_n^h - E_n^c$, $\Delta p_n = p_n^h - p_n^c$ and $W < 0$ indicates work performed by the system. The efficiency of the cycle
is

\begin{equation}
\eta = -\frac{W}{Q_h}.
\end{equation}

\section{Engines based on Unital measurements}

Let us now consider a measurement-based Otto engine, as proposed by Yi et al. \cite{yi2017single}, where the interaction with the hot bath in the third stroke above is replaced by a general trace-preserving measurement process $\mathcal{E}$, characterized by Kraus operators
$\{M_\alpha\}$ satisfying $\sum_{\alpha} M_{\alpha}^{\dagger}M_{\alpha} = \mathds{1}$. The density operator after the measurement is 
\begin{equation}
\rho_M = \mathcal{E}(\rho(\lambda_{f})) = \sum_{\alpha} M_{\alpha} \rho(\lambda_f) M_\alpha^\dagger.
\end{equation}
As mentioned before, the quantum-mechanical nature of the measurement will generally disturb the system, and in particular its energy, in random and irreversible ways. Therefore various authors have argued that any energy transferred in this process should also be interpreted as a form of heat \cite{elouard2017extracting,jordan2020quantum}. 

%Note also that this process only changes the population of the density matrix, as in the usual definition of heat for a quantum system. \cite{alguem}

%%% Nota DJ: Não entendi essa afirmação. Uma medida geral pode também afetar as coerencias da matriz densidade.

Since we want the measurement to play the role of the hot bath, we must ensure that, on average, it will increase the system's energy, i.e., that
\begin{align}
\left\langle \Delta E \right\rangle = \text{Tr} \left[(\rho_{M} - \rho(\lambda_{f}) )\mathcal{H}(\lambda_{f})\right] \geq 0. \label{eq:hotmeasure}
\end{align}

In Ref. \cite{yi2017single}, it was shown that when $\rho(\lambda_{f})$ is a passive state (as is the case here), Eq.\,(\ref{eq:hotmeasure}) is indeed always satisfied for `minimally disturbing measurements' (MDM's), namely those where $M_\alpha=M_\alpha^\dagger$ are all Hermitian \cite{wiseman2009book}.
This choice seems, however, to have been motivated more by mathematical convenience than by physical considerations. From a physical standpoint, it is more natural to consider the full set of \emph{unital} channels, those that map the identity operator to itself, i.e, satisfy $\sum_{\alpha} M_{\alpha}M_{\alpha}^{\dagger} = \mathds{1}$, since these channels increase the von Neumann entropy $S = -\text{Tr} \rho \ln \rho$ for all input states,  $S(\mathcal{E}(\rho)) \geq S(\rho)$ \cite{holevo2019book}. Note that the MDM's considered in Ref.  \cite{yi2017single} are a special case within this class.

%But actually it easy to see that this
%is also true for unital measures, which also have the property to
%increase the entropy of the state. 

%%% Nota DJ: NO IT ISN'T EASY... tentei explicar isso diversas vezes mas pelo visto sem sucesso. Isso tem de ser mostrado separadamente, pois o argumento usado em Yi et al não se aplica no caso de um unital geral.

In fact, we can show that Eq.\,(\ref{eq:hotmeasure}) remains true for arbitrary unital channels:
\smallskip

\textbf{Theorem 1}: If $\mathcal{E}$ is a unital, completely positive quantum channel and $\rho$ is a passive quantum state with respect to Hamiltonian $\mathcal{H}$, then $\text{Tr} \left[ (\mathcal{E}(\rho)-\rho) \mathcal{H} \right]\geq0$.\smallskip

Since this statement is possibly of more general interest, and requires a somewhat more general argument than the one presented in Ref.  \cite{yi2017single}, we give a full proof in Appendix \ref{sec:appendA}.
\smallskip

%Thus, unital measurements seem to be the most general class that can act as a hot bath giving energy as heat to the system

Let us now consider the heat and work exchanges for this measurement-based engine.
The only difference from the conventional one is that after the third stroke the state of the system is $\rho_M$ instead of the thermal state $\rho(\lambda_f)$ at inverse temperature $\beta_h$. Thus
Eqs.\, (\ref{eq:Qc}) and (\ref{eq:Qh}) can be rewritten as
\begin{align}
Q_c^{M} &= \sum_n E_n(\lambda_i)[p_n^c - p_n^{M}] \leq 0 \\
Q_h^{M} &= \sum_n E_n(\lambda_f)[p_n^{M} - p_n^c] \geq 0
\end{align}
where
\begin{equation}
p_n^M  \equiv  \langle E_n(\lambda_f)|\rho_{M}| E_n(\lambda_f) \rangle
\end{equation}
are the populations of the energy basis states after the measurement  \footnote{As discussed in Appendix \ref{sec:appendA}, it is possible to write $p_n^M =  \sum_m p_m(\lambda_i)T_{nm}$
with $T_{nm}  \equiv \sum_\alpha | \langle E_n(\lambda_f)|M_\alpha| E_m(\lambda_f) \rangle|^2$ forming a bistochastic transfer matrix}.
Similarly the work is given by
\begin{equation}
\begin{split}
W &=  -\sum_n[E_n(\lambda_f) - E_n(\lambda_i)][p_n^{M} - p_n^c].
\end{split}
\end{equation}
The engine efficiency is therefore
\begin{align} \label{eq:Ottoefficiency}
\eta &= \frac{\sum_n[E_n(\lambda_f) - E_n(\lambda_i)][p_n^{M} - p_n^c].}{\sum_n E_n(\lambda_f)[p_n^{M} - p_n^c] }
\end{align}

All these expressions are identical to those for a two-bath  engine, apart from the correspondence between $p_n^M$ and $p_n^h$.  In particular, consider a scenario where all the energy gaps change by the same ratio $r$ when $\lambda$ is adiabatically increased, i.e, 
\begin{align} \label{eq:gapratios}
E_n(\lambda_f)-E_{m}(\lambda_{f}) = r [E_n(\lambda_i)-E_{m}(\lambda_{i})], \, \forall{n,m}.
\end{align}
 In this case, as has already been shown both for the two-bath case \cite{quan2007quantum} and for the measurement-based one \cite{yi2017single}, $\eta$ does not depend on the probability changes $\Delta p_n$, but only on $r$:
\begin{align} 
\eta =\frac{r-1}{r} = 1 - \frac{1}{r} \equiv \eta_{0}.  \label{eq:ordinaryOtto}
\end{align}
In these cases it does not matter if it the heat comes from a thermal interaction or a measurement process, since the efficiency does not depend on the temperatures of the reservoirs, nor on the choice of measurement.
One way to understand this correspondence is to note that, when Eq.(\ref{eq:gapratios}) holds, the system's state does retain a thermal form during the adiabatic strokes, with an effective temperature that depends on $\lambda$. Thus, here the thermodynamic and quantum notions of an adiabatic process do coincide \cite{quan2007quantum}.

This, however, raises the following question: are there circumstances where we can  improve the efficiency of a quantum Otto engine beyond $\eta_{0}$? The previous discussion shows this requires at least one level gap that does not adiabatically shift with the same ratio as the others. One way to realize this possibility is by using a working substance composed of interacting spin-1/2 particles, as studied for the two-bath model in Ref. \cite{thomas2011coupled} and for the measurement-based model in Ref. \cite{das2019measurement}. Improvements in efficiency were found in both works. However no clear physical explanation for this effect was given.

More recently, after analyzing this same system for the two-bath case, some of us have found that it does possess a simple mechanism for efficiency increase \cite{Oliveira20}. This mechanism does not depend on any correlations, but on the exploitation of reversed heat fluxes (from the cold bath to the hot) via uncoupled levels. In the next section we show that the same mechanism can also be present in a measurement-based engine, and can in fact deliver an even greater improvement in efficiency. 
%We show this using a toy model `working substance' with three energy levels, one of which remains fixed during the adiabatic process, ensuring that not all levels change by the same ratio. 

\section{QUTRIT as a working substance}

A two-level system (qubit) only has a single energy gap, so Eq.(\ref{eq:gapratios}) is trivially satisfied. In other words, the simplest possible Otto engine where we can hope to see an increase in $\eta$ beyond $\eta_{0}$ has a three-level (or qutrit) working substance. In the following we consider a qutrit governed by the Hamiltonian:
\begin{equation}
\mathcal{H} = 
\left[
\begin{matrix} 
0  &  B  &  0  \\
B  &  0  &  0  \\
0  &  0  & -J 
\end{matrix}
\right],
\end{equation}
where $B>0$ plays the role of the adiabatic parameter $\lambda$, shifting between values $B_{i}$ and $B_{f}$, while $J>0$ is kept fixed. This toy model can be considered as a simplification of the coupled-qubit system studied in \cite{thomas2011coupled, Oliveira20}. Physically it could, for instance, be realised in a three-level atom with a $V$-type level structure, with two initially degenerate upper levels Raman-coupled via the lower one. The eigenvalues and eigenstates are
%As mentioned before, this is a simple toy-model but it will be sufficient to understand the mechanism of an increase or decrease in the efficiency due to the coupling and/or the measurement process.
\begin{align}
\begin{tabular}{  c | c } 
Eigenvalues & Eigenstates  \\ 
\hline
$+B$ & $\ket{+}= (\ket{0} + \ket{1})/ \sqrt{2}$ \\
\hline
$-B$ & $\ket{-} = (\ket{1} - \ket{0})/ \sqrt{2}$  \\
\hline
$-J$ & $\ket{2}$\\
\hline
\end{tabular}
\end{align}
%\begin{center}
%\begin{table}[!ht]
%\begin{tabular}{ | c | c | } 
%\hline
%Eigenvalues & Eigenstates  \\ 
%\hline
%$+B$ & $\ket{+}= (\ket{0} + \ket{1})/ \sqrt{2}$ \\
%\hline
%$-B$ & $\ket{-} = (\ket{1} - \ket{0})/ \sqrt{2}$  \\
%\hline
%$-J$ & $\ket{2}$\\
%\hline
%\end{tabular}
%\caption{Eigenvalues and eigenstates for a qutrit system given by the Hamiltonian (14)}
%\end{table}
%\end{center}

The resulting heat exchanges are given by
\begin{align}
 Q_h  &=  -J (\Delta p_{-J}) + B_f (\Delta p_{B}) -B_f (\Delta p_{-B}),  \label{eq:Qhqutrit}\\
 Q_c &=  +J (\Delta p_{-J}) - B_i (\Delta p_{B}) +B_i (\Delta p_{-B}), \label{eq:Qcqutrit} 
\end{align}
and the work by
\begin{equation}
W = (B_f-B_i) (\Delta p_{B} - \Delta p_{-B}), 
\end{equation}
where $\Delta p_n=p_n^h-p_n^c$ for the two-bath model, $\Delta p_n=p_n^M-p_n^c$ for the measurement model, and we have labeled each probability by their corresponding energy.   

Using terminology introduced in Ref. \cite{Oliveira20}, the energy levels $\pm B$ are `working' levels, since they shift with the adiabatic parameter. Level $-J$, which does not shift, is `idle'.  As shown more generally in Ref. \cite{Oliveira20}, the presence of idle levels allows one to either increase or decrease the efficiency of an Otto cycle away from $\eta_{0}$. Indeed, in the current example, the efficiency is
\begin{equation}
 \frac{\eta}{\eta_0} = 1 + J\frac{\Delta p_{-J}}{Q_h}.
\label{qutriteff}
\end{equation}
We should mention that, while normally the time scale for an adiabatic process increases with the inverse of the smallest energy gap, here this is not necessary, since a time-dependent change $B(t)$ introduces no crosstalk between levels. Thus, in fact, the `adiabatic' strokes can be executed here at finite speed \cite{Oliveira20}.

To understand Eq. (\ref{qutriteff}), it is convenient to interpret each term in Eqs. (\ref{eq:Qhqutrit}, \ref{eq:Qcqutrit})  as a separate energy flux, e.g. to view $-J\Delta p_{-J}  \equiv q_{-J}^{h}$ as the heat absorbed by the engine from the hot bath \emph{via} the  $-J$ energy level.  In terms of this quantity, we can write
\begin{equation}
 \frac{\eta}{\eta_0} = 1 - \dfrac{q_{-J}^{h}}{Q_h}.
\label{ratioeff}
\end{equation}

Considering that an engine requires $Q_{h} >0$, we can now see that an increase in efficiency is only possible if $q_{-J}^{h}< 0$ (or, equivalently, if $\Delta p_{-J} >0$). Note also that, since level $-J$ is idle, any heat it absorbs from one bath cannot be converted into work, but must be deposited in the other, hence $q_{-J}^{c} = - q_{-J}^{h}$. In other words, although the overall heat flow in an engine cycle is from the hot bath to the cold, attaining efficiency greater than $\eta_{0}$ requires part of the heat to flow in the \emph{opposite} direction, via the idle level $-J$.  It turns out that such a reversed heat flow is indeed possible in many situations \cite{Oliveira20}.

Note that the above conclusion holds both for the measurement-based engine and in the two-bath scenario. As already stressed, the only difference between the two situations is the origin of the
populations after the third stroke. In the two-bath model, for a given pair of field values $B_{i}, B_{f}$, these populations depend only on the hot bath temperature.  In the measurement-based model, however, they depend on the choice of measurement, which has many more free parameters. One then expects that an appropriate choice may lead to an increase in $\Delta p_{-J}$ beyond what is possible with a thermal bath - and thus to a higher efficiency. In the following, we show that this is indeed the case.

\subsection{Two-bath Model}
We begin by writing explicit expressions for the heat and work exchanges in the case of the two-bath model: 
\begin{widetext}
\begin{align}
Q_h = \dfrac{(2e^{-\beta_h B_f} + e^{\beta_h J}) B_f -J e^{\beta_h J}}{2\cosh{(\beta_hB_f)} + e^{\beta_hJ}} - \dfrac{(2e^{-\beta_c B_i} + e^{\beta_c J}) B_f -J e^{\beta_c J}}{2\cosh{(\beta_cB_i)} + e^{\beta_cJ}},
\\
 Q_c =  -\dfrac{(2e^{-\beta_h B_f} + e^{\beta_h J}) B_i -J e^{\beta_h J}}{2\cosh{(\beta_hB_f)} + e^{\beta_hJ}} + \dfrac{(2e^{-\beta_c B_i} + e^{\beta_c J}) B_i -J e^{\beta_c J}}{2\cosh{(\beta_cB_i)} + e^{\beta_cJ}},
\end{align}
\begin{align}
W = (B_f - B_i)\Big(\dfrac{2e^{-\beta_h B_f} + e^{\beta_h J}}{2\cosh{(\beta_hB_f)} + e^{\beta_hJ}} - \dfrac{2e^{-\beta_c B_i} + e^{\beta_c J}}{2\cosh{(\beta_cB_i)} + e^{\beta_cJ}}\Big).
\end{align}

The resulting efficiency can be written as
\begin{equation}
\eta = \frac{(B_f - B_i)}{B_f + \Omega J} = \frac{\eta_0}{1 +(\Omega/B_f) J},
\end{equation}
with
\begin{equation}
\Omega = \frac{e^{\beta_c(B_i+J)}+e^{\beta_c(B_i+J)+2\beta_hB_f} -e^{\beta_h(B_f+J)+2\beta_cB_i}-e^{\beta_h(B_f+J)}}{2(e^{2\beta_cB_i} - e^{2\beta_hB_f}) + e^{\beta_c(B_i+J)} - e^{\beta_h(B_f+J)} - e^{\beta_c(B_i+J)+2\beta_hB_f} +  e^{\beta_h(B_f+J)+2\beta_cB_i}}.
\end{equation}
\end{widetext}

As expected, when $J=0$, we recover $\eta = \eta_0=1-B_i/B_f$, the same result that is valid for a two-level system.
Note that while $\eta_0$ does not depend explicitly on the bath temperatures, in order to operate as an engine, $W<0$, we need $T_h \geq (B_f/B_i) T_c$. 

Our main interest is to compare the effects of having $J\neq 0$ on the efficiency of the two-bath and measurement engines. Unfortunately, for the qutrit there is no simple general expression for the condition where the cycle operates as an engine; this condition occurs only in the limit of high and low temperatures, as shown in \cite{Oliveira20} for the analogous two-qubit system.  Thus, in order to keep the problem tractable, in the remainder of this article we set the parameters $B_i = 3$, $B_f = 4$, $\beta_h = 0.5$ and $\beta_c = 1$, and focus on analyzing the effects of changes in $J$ and in the measurement protocol. 

\begin{figure}[h]
\center
\includegraphics[scale=0.33]{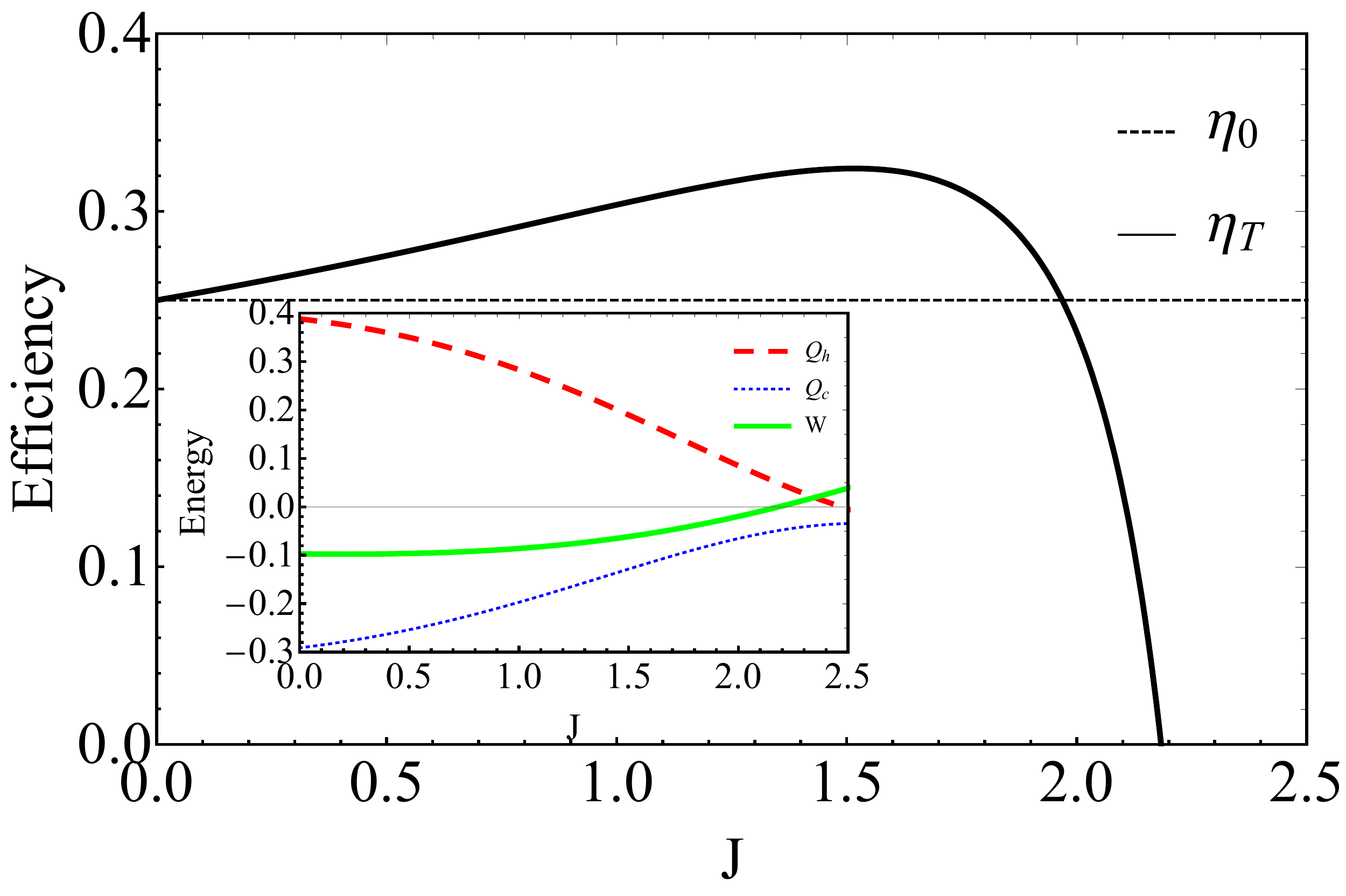}
\caption{Efficiency of a two-bath quantum heat engine with a qutrit `working substance'. $\eta_0$ is the efficiency for a two-level case ($J = 0$; dashed black). $\eta_T$ is the efficiency for the qutrit (solid black). (Inset) Heat absorbed by the system (top dashed red), heat releasedon the system (middle solid green). Note that, in our sign convention, negative $W$ means work is being extracted, i.e, the system is functioning as a heat engine.}
\label{Fig-Eff-Bath}
\end{figure}

For these parameters, we first plot in Fig.\ref{Fig-Eff-Bath} the efficiency as a function of $J$ for the qutrit `two-bath' model. We can see that $\eta$ at first increases with
$J$, reaches a maximum, and then decreases, becoming smaller than $\eta_0$ 
and finally becoming negative; the system stops to operate as an engine since $W>0$.
In the inset of Fig.\ref{Fig-Eff-Bath}, we plot both heats and work. Note that while $\eta$ increases with $J$, the amount of work delivered per cycle decreases, since more energy is flowing through the $-J$ level and cannot be converted into work.
This behavior is similar to that obtained when using two spins coupled via the Heisenberg interaction \cite{thomas2011coupled,Oliveira20}, since the effect
of the coupling is to introduce an idle level.

\normalsize

\subsection{Measurement-based model}

Let us now consider a measurement-based version of the Otto engine for the qutrit. We restrict ourselves to projective (von Neumann) measurements and choose one possible set of projection operators given by:
\begin{equation}
M_1 = \ket{\psi_1} \bra{\psi_1},
M_2 = \ket{\psi_2} \bra{\psi_2},
M_3 =  \ket{\psi_3} \bra{\psi_3},
\end{equation}
where
\begin{align*}
\ket{\psi_1} &= \cos{\theta} \sin{\phi} e^{i\chi}\ket{0} + \sin{\theta} \sin{\phi} e^{i\psi}\ket{1} + \cos{\phi}\ket{2} \\ 
\ket{\psi_2} &= \cos{\theta} \cos{\phi} e^{i\chi}\ket{0} + \sin{\theta} \cos{\phi} e^{i\psi}\ket{1} - \sin{\phi}\ket{2} \\
\ket{\psi_3} &= \sin{\theta} e^{i\chi}\ket{0} - \cos{\theta} e^{i\psi}\ket{1}
\end{align*}
This set of projectors belongs to the $SU(3)$ group and can be used to make a generic von Neumann measurement in any direction. In this case we do not have only one parameter, $\beta_h$, but instead four:  $\theta, \phi, \chi$ and $\psi$. The expressions for the heat, work and efficiency can be calculated but become very cumbersome, so we do not present them here.

\begin{figure}[!ht]
\center
\includegraphics[scale=0.35]{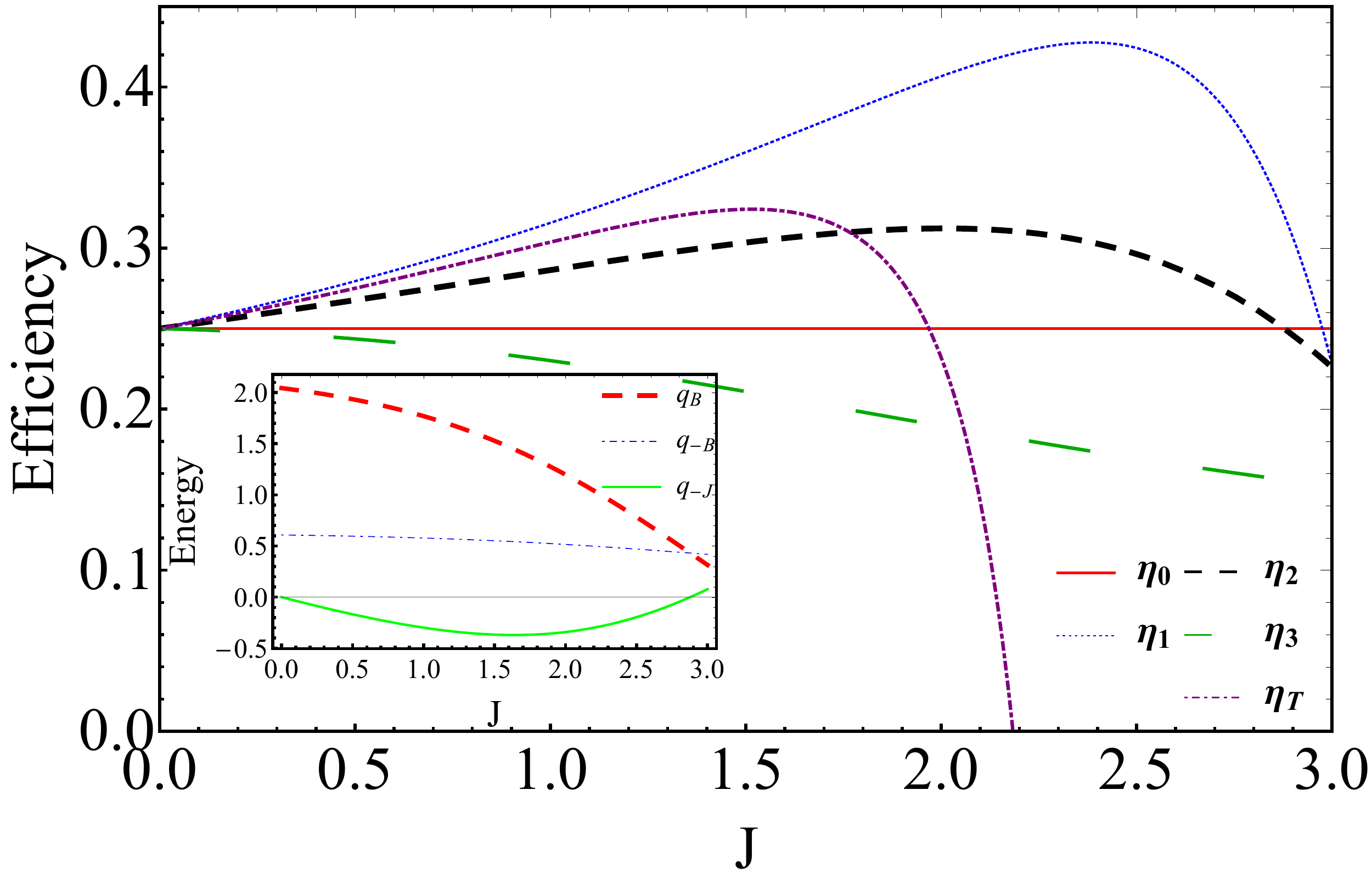}\\
\caption{Efficiency of the measurement-based quantum heat engine for a qutrit. $\eta_1$ (first curve, dotted blue): $\theta=\phi=0.7\pi$ and $\psi=\chi=0.5\pi$; $\eta_2$ (third curve, dashed black):  $\theta=\phi=\chi= 0.7\pi$ and $\psi=0.5\pi$; $\eta_3$ (fifth curve, dashed green): $\theta=\phi=\chi=\psi=0.3\pi$. We also show the efficiency for the two-bath model with the qutrit $\eta_T$ (second curve, dotdashed purple) and with the qubit $\eta_0$ (fourth curve, solid red). (Inset) Energy exchanged through each energy level for $\eta_2$: $q_B$ (top dashed red), $q_{-B}$ (middle dotdashed blue), $q_{-J}$ (bottom solid green).}
\label{Fig-Eff-Meas}
\end{figure}

The analytical expressions for the efficiency are also very cumbersome and do not give much insight. Thus we have numerically analyzed the efficiency for many different values of $\theta, \phi, \chi$ and $\psi$, and we will now show the more interesting and representative results. In Fig.\,\ref{Fig-Eff-Meas} we show
the efficiency for three different measurements, for the two-baths
model and for the qubit (which is the same for the measurement or two-bath model). It can be seen that the measurement-based model can have higher or lower efficiency than the two-bath model and even lower efficiency than the qubit system.

It is important to emphasize that in both engine models, two-baths and measurement-based, the increase in the efficiency in relation to the qubit is due to the flow of energy in the $-J$ idle level being from the cold to the hot bath. This can be seen in the inset of Fig.\ref{Fig-Eff-Meas} for one of the measurement protocols. The fact that the efficiency of the measurement model can be larger than that of the two-bath model is due to the measurement being able to give more energy to the $-J$ level than the thermal hot bath. We also checked that for the same set of parameters from Fig.\ref{Fig-Eff-Meas}, the efficiency of the two-bath engine always decreases for negative values of $J$, while it can increase for the measurement-based engine.
Note also that in both models, the increase in the efficiency is not related
to entanglement or any other subsystem correlations, which are not present in
a single system with three levels.

\begin{figure}
\centering
\subfigure[]{
	\includegraphics[width=4.1cm,height=4.5cm]{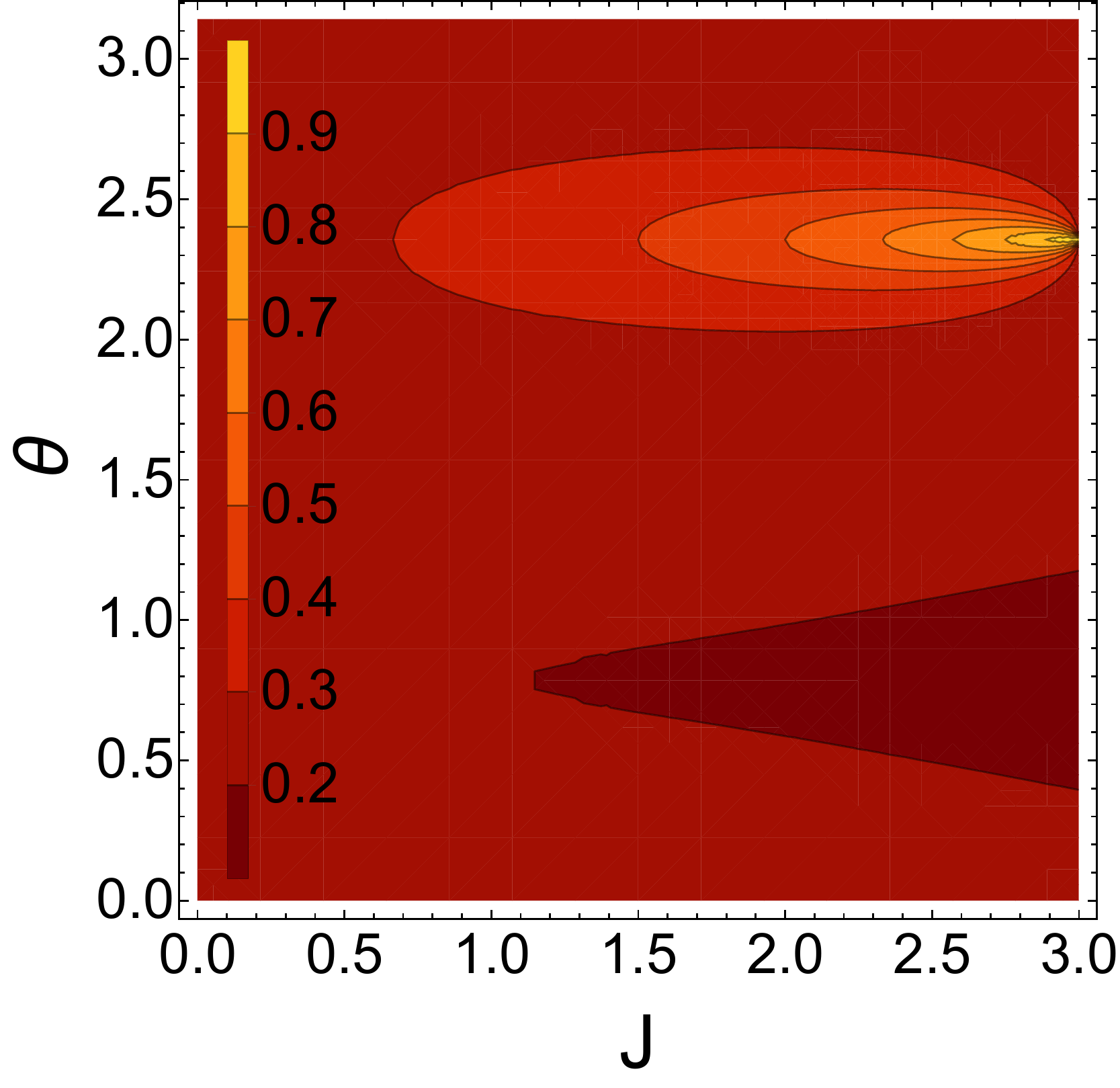}%
}
\subfigure[]{
     \includegraphics[width=4.1cm,height=4.5cm]{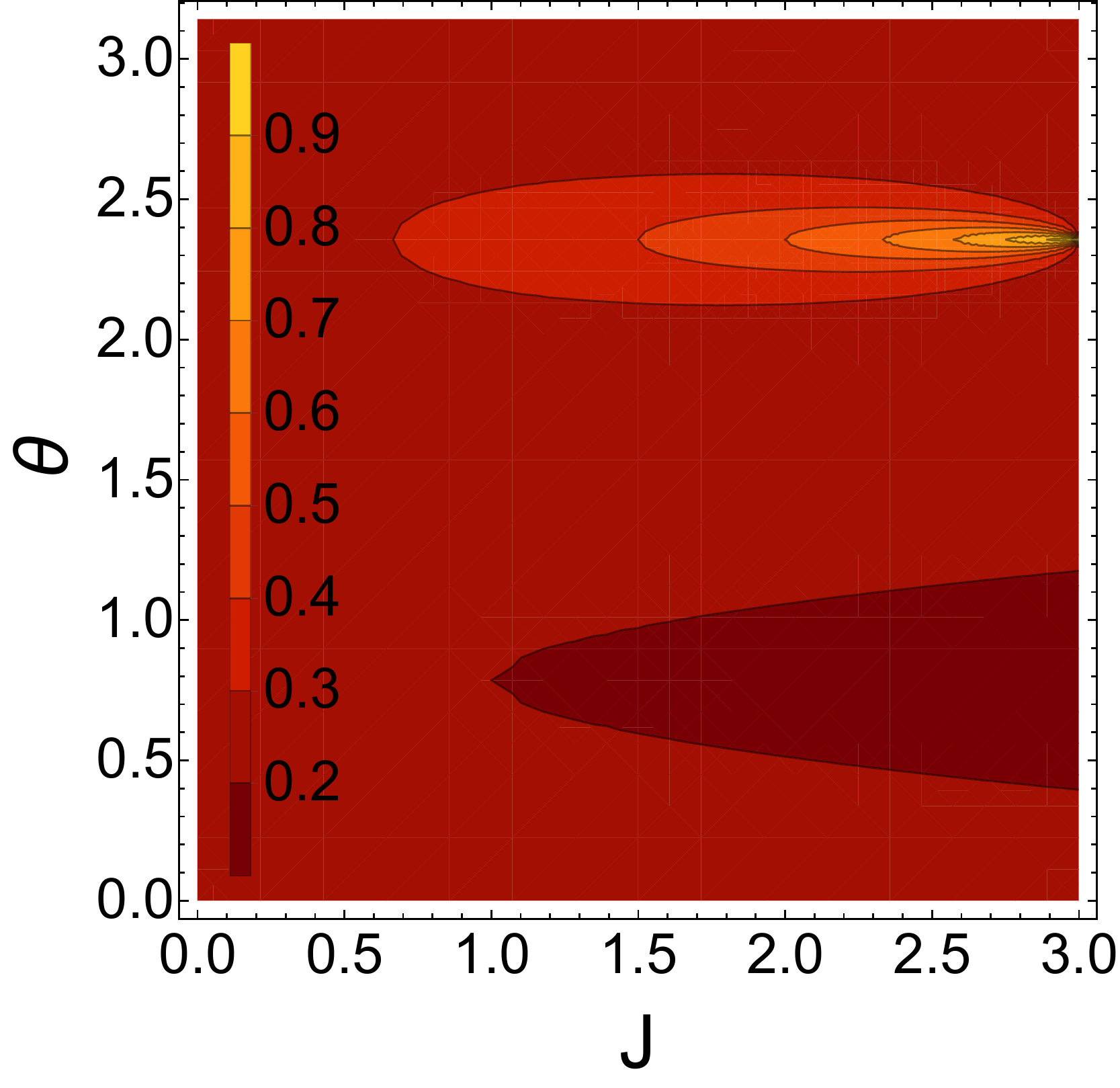}%
}
\subfigure[]{
     \includegraphics[width=4.1cm,height=4.5cm]{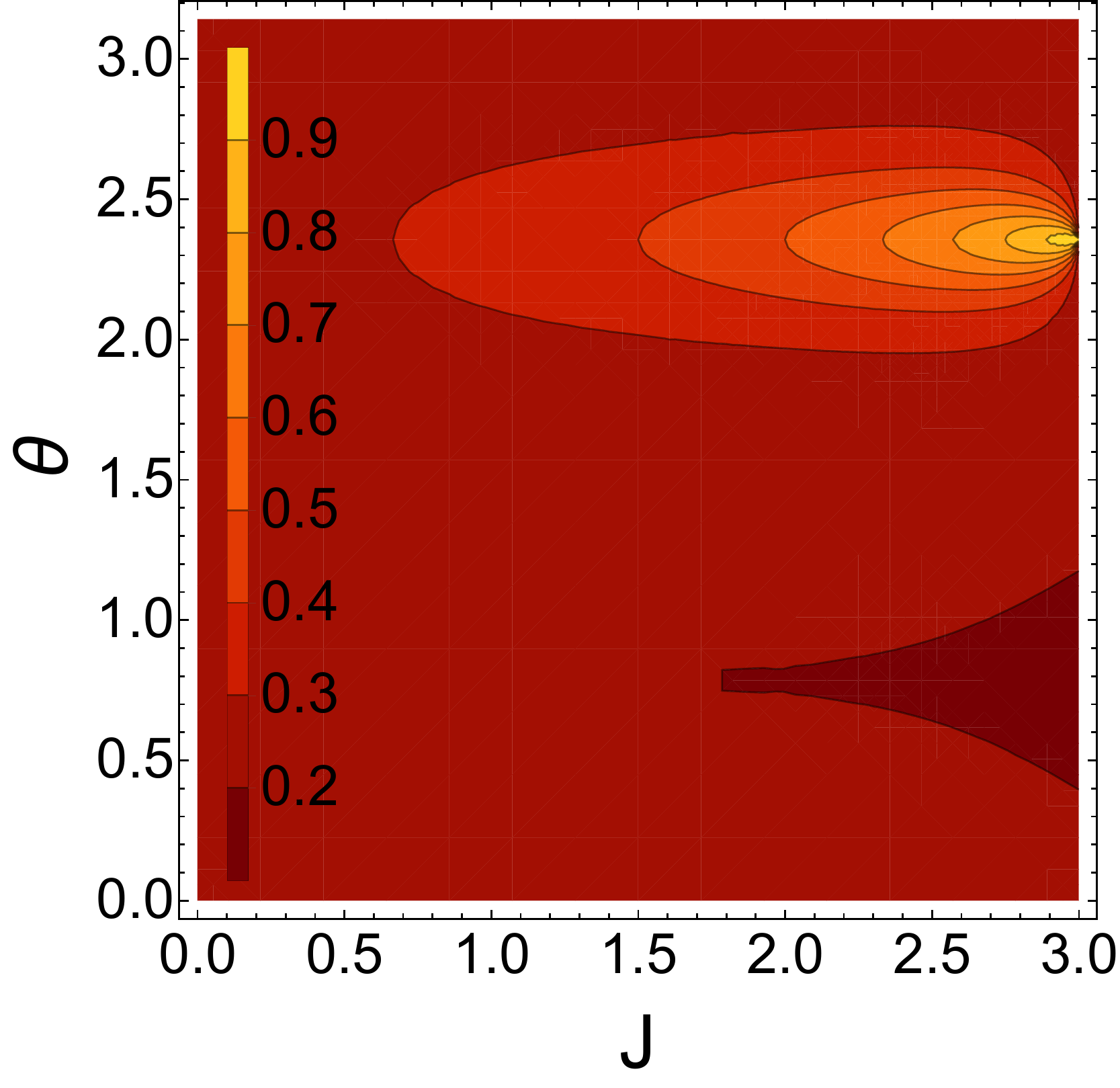}%
}
\subfigure[]{
     \includegraphics[width=4.1cm,height=4.5cm]{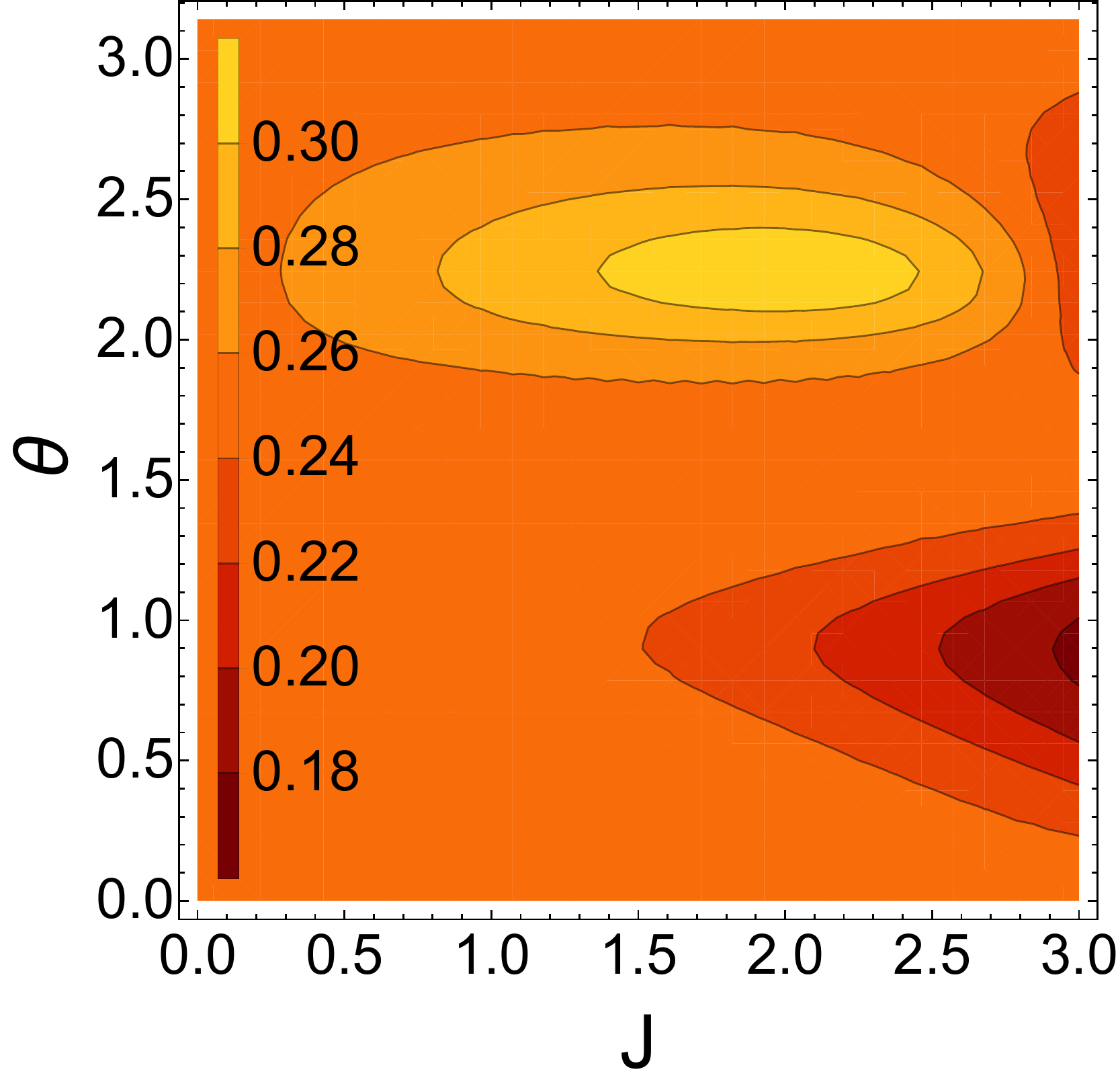}%
}
\caption{Contour plot of the efficiency of the measurement-based quantum heat engine for a qutrit. In figs (a), (b) and (c) $\theta=\phi$ and $\psi = \chi = \pi/2$
and $\beta=1, 0.25, 4$ for (a), (b) and (c). In fig. (d)
$\theta = \phi = \chi$, $\psi=\pi/2$ and $\beta_c=1$.}
\label{curvacontor}
\end{figure}

We also numerically found that the efficiency of the measurement-based engine
can approach $1$. This
can be seen in Fig. \ref{curvacontor}(a) where we have a contour plot
of the efficiency as a function of $\theta = \phi$ and $J$ for 
$\psi = \chi =\pi/2$ and $\beta_c=1$. One can see that for some
fixed value of $\theta = \phi$ around $2.4$ the efficiency increases
with $J$ and seems to approach $1$ as $J\rightarrow 3$. 
In figs. \ref{curvacontor}(b) and (c) we analyze the effects of changing
the cold bath temperature. It can be seen that lower (higher) $\beta_c$ decreases (increases) the size of the region of higher efficiency around the maximum,
which can reach $1$. Finally we show in Fig. \ref{curvacontor}(d) the
case where $\theta = \phi = \chi$ and $\psi=\pi/2$ and $\beta_c=1$.
It can be seen that the efficiency does not reach $1$ anymore, and
the maximum values occur around $J=2$.

\begin{figure}
\centering
\subfigure[]{
     \includegraphics[width=4cm,height=4cm]{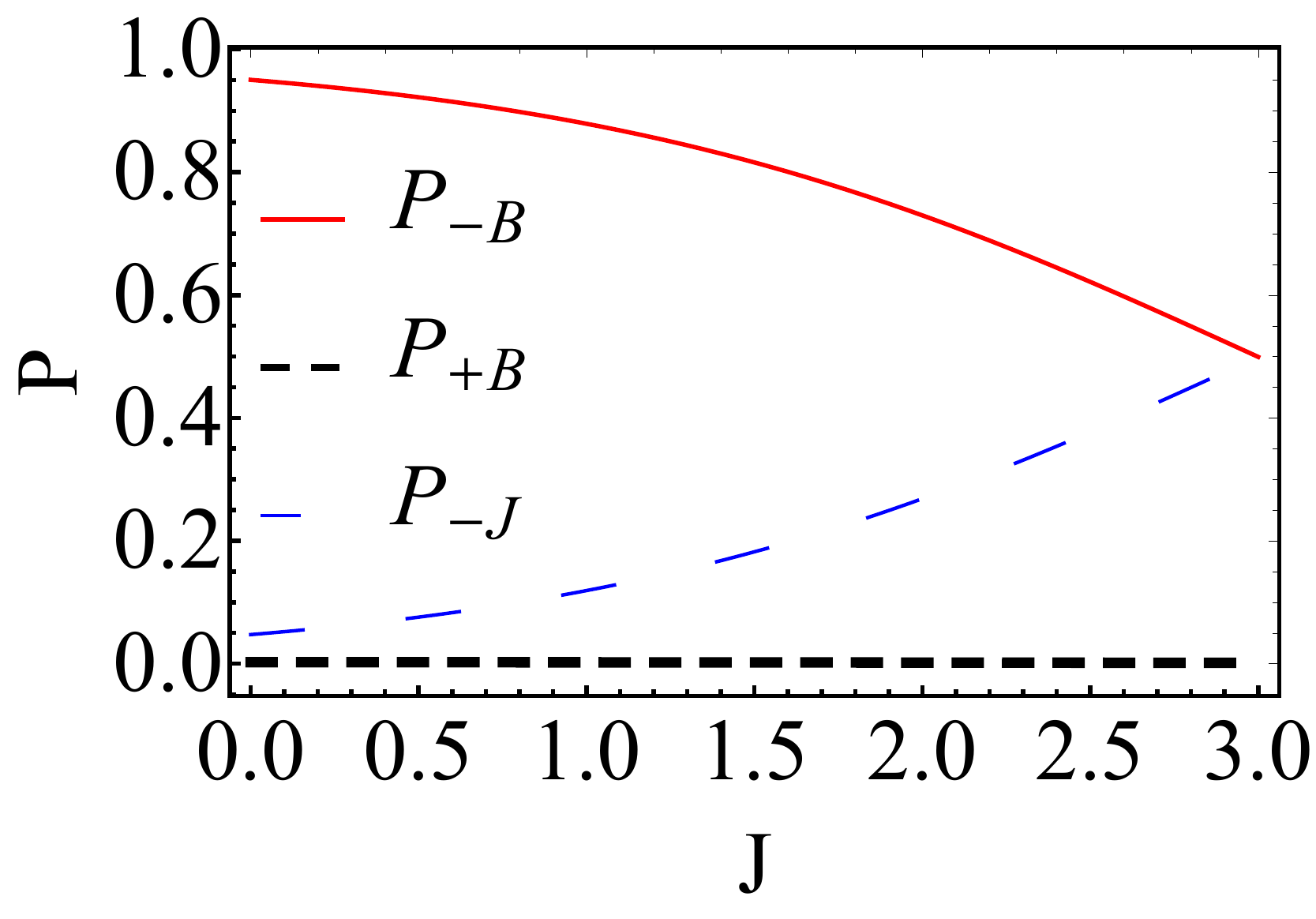}%
}
\subfigure[]{
     \includegraphics[width=4cm,height=4cm]{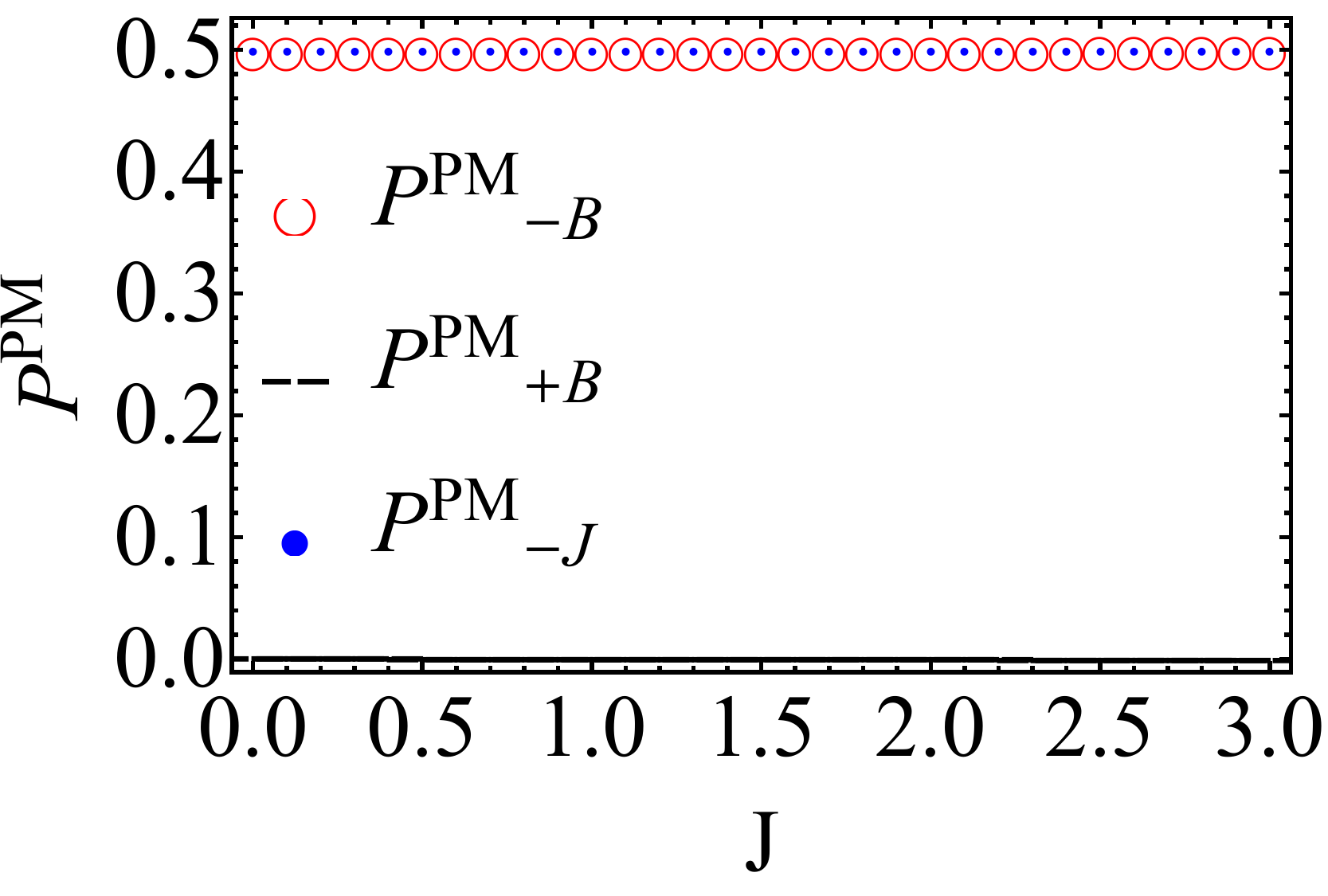}%
}
\subfigure[]{
     \includegraphics[width=4cm,height=4cm]{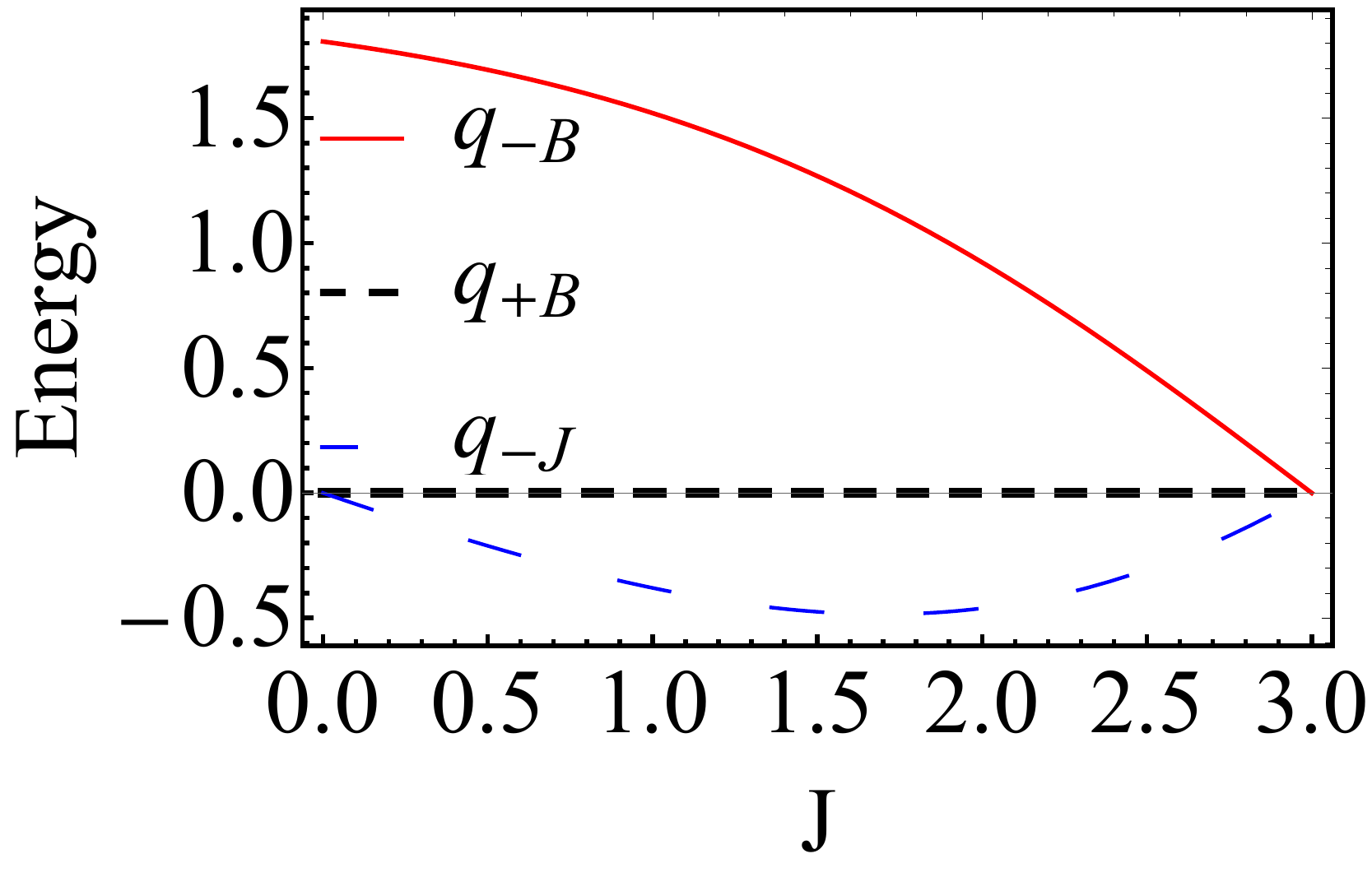}%
}
\subfigure[]{
     \includegraphics[width=4.3cm,height=4cm]{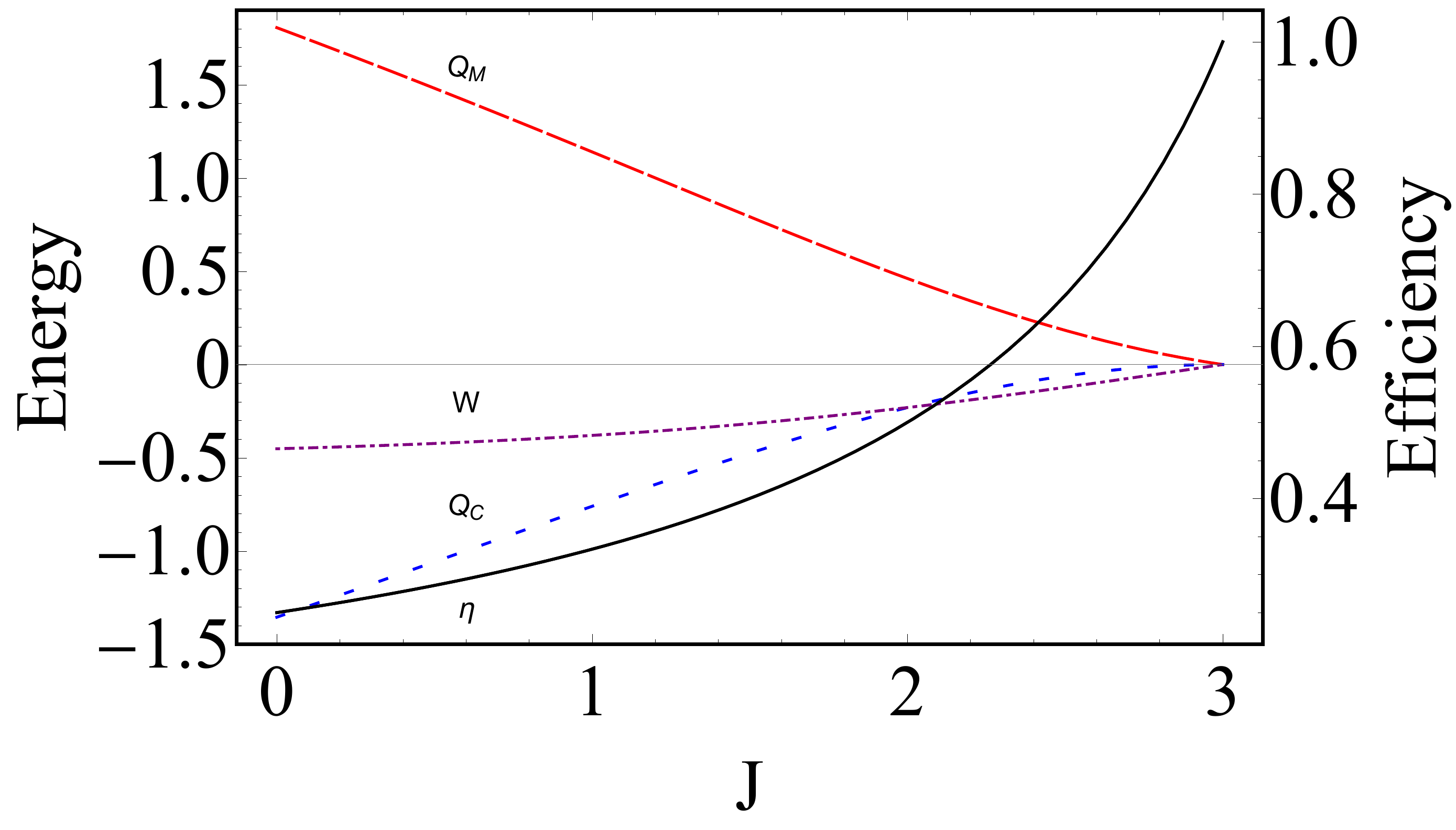}%
}
\caption{Energy and population structure for the qutrit system: (a) Initial (cold) thermal population for $\beta=1$. (b) Population after the measurement process, third stroke of the cycle. (c) Energy flowing through each energy level during the third stroke. (d) Efficiency (solid black), total work (dotdashed purple), absorbed heat (dashed red) and released heat (dashed blue) during the cycle. All this graphics were done for the angles $\theta=\phi=0.75\pi$ and $\chi=\psi=0.5\pi$}
\label{fig-pop}
\end{figure}

We now analyze more carefully the extreme scenario where
the efficiency can approach $1$ in Fig. \ref{curvacontor} (a). 
We fixed  $\theta=\phi=0.75\pi$ and looked at the behavior
of the efficiency, populations and exchanged heat via the idle level $J$. 
In Fig. \ref{fig-pop}(a) we show the populations of the
system state in the energy basis before the measurement. First
note that the population of the highest level, $+B$, is negligible.
We also see that as we increase $J$, the populations of the two lowest
levels approach each other, as expected since the gap between the 
two levels is decreasing. In Fig.  \ref{fig-pop}(b) we show the
populations after the measurement. One can see that for all values
of $J$, the measurement does not change the population of the highest
level, $+B$, and it projects the state in an equal mixture of the two lowest
energy levels. Comparing the population before and after the 
measurement, we see that as $J$ increases, the change in the population
of the two lowest levels decreases and goes to $0$ when the
efficiency approaches $1$; so less energy is transferred to the system by the measurement.
This can be seen in Fig. \ref{fig-pop}(c), where we plot the energy exchanged
by each level during the measurement; they all go to $0$ in the limit where
the efficiency approaches $1$. Finally, in Fig. \ref{fig-pop}(d) we show the heats and the total work: they all go to $0$ in the limit of efficiency $1$. Thus while we have found measurement protocols that can reach high values of efficiency and even approach $1$, the work produced by the engine decreases and ultimately reaches $0$: we have a very efficient engine, but it produces negligible work. As mentioned, in this limit the effects of the measurement also become negligible.

It is worth noting that the Carnot limit does not apply to our engine, since one of the baths is not a thermal bath. There are also other examples of engines with non-thermal baths, for example one consisting of  squeezed bath instead of thermal bath, that may surpass the Carnot limit; there are other upper
bounds for such engines, but it is not clear if they apply here \cite{niedenzu2018quantum}.

In summary, we have shown that the mechanism for an increase in the efficiency when we add an energy level that does not change during the adiabatic process (an ``idle" level) is the same for the two-bath and the measurement model. The increase is due to a flux of energy from the cold to the hot bath through the ``idle" level. We have also shown that for a qutrit the 
measurement-based engine can have higher efficiency than the analogous two-bath model, due to the greater range of possible changes in the populations that may be induced through a measurement. Finally it is even possible to reach efficiencies close to $1$, but the output work becomes negligible.

\section{Revisiting the Heisenberg Model as a QHE}

Another possibility to study quantum engines beyond the qubit case is to couple two qubits. In fact, coupled spin-1/2 models have been extensively studied as thermal engines. In these models one may expect that the quantum correlation between the spins may enhance engine efficiency. However there is no clear connection between any measure of quantum
correlation and efficiency increase. Here we will show that the same mechanism of efficiency increase for a qutrit can explain the results obtained for two spins coupled via a Heisenberg Hamiltonian ($XXZ$ model), which
were already analyzed for the two bath model \cite{thomas2011coupled} and the measurement model \cite{das2019measurement}. Thus in this model correlations
are not needed to explain the efficiency increase.

%We believe that the author of the first one considered a misleading interpretation of the functions $q_n$. In the second paper, it was not clear the mechanism that enhanced the efficiency of the engine. Here we show that we can explain this boost in the efficiency using the same mechanism presented in the previous section.

The anisotropic Heisenberg Hamiltonian for two spins
1/2 ($XXZ$ model) is given by
\begin{equation}
\mathcal{H} = J_{xy}(\sigma^1_x\sigma^2_x + \sigma^1_y\sigma^2_y) + J_z\sigma^1_z\sigma^2_z + B(\sigma^1_z+\sigma^2_z),
\end{equation}
with $J_{xy}$ the interaction constant of the spins in the $xy$-plane, $J_z$ the interaction constant in the $z$-direction, $B$ the external magnetic field in the $z$-direction, and $\sigma^{1(2)}_i$ the Pauli matrices associated with the particle $1(2)$. The eigenvalues and eigenvectors are given in Table \ref{heisenbergeigen}.

\begin{table}[H]
\centering
\begin{tabular}{ |c|c|c|c|c| } 
\hline
Eigenvalues & Eigenstates  \\ 
\hline
$2B$ & $\ket{00}$  \\
\hline
$2(J_{xy} - J_z)$ & $(\ket{01} + \ket{10})/ \sqrt{2}$ \\
\hline
$-2(J_{xy} + J_z)$ & $-(\ket{01} - \ket{01})/ \sqrt{2}$ \\
\hline
$-2B$ & $\ket{11}$ \\
\hline
\end{tabular}
\caption{The four eigenvalues of the Hamiltonian $\mathcal{H}$ with their associated eigenvectors.}
\label{heisenbergeigen}
\end{table}

As in the qutrit case, there are ``idle" energy levels that
do not depend on the external parameter, $B$, and
therefore do not contribute to the total work. In this case
there are two ``idle" levels, but the same manipulations
that lead to Eq.\,(\ref{ratioeff}) can be made and the efficiency can be written as
\begin{equation}
\dfrac{\eta}{\eta_0} = 1 - \dfrac{q_1 + q_2}{Q_{h}},
\label{heiratio}
\end{equation}
where $\eta_0=1-B_i/B_f$ is the efficiency for $J_{xy}=J_z=0$ and
$q_1$ and $q_2$ are the energy exchanged through the two ``idle" energy levels: $2(J_{xy}-J_z)$ and $-2(J_{xy}+J_z)$. 
So the condition for the coupling to increase efficiency is $q_1 + q_2 < 0$. 
As before, this means that, the total energy flowing through the two ``idle" energy levels
has to be from the cold to the hot bath. As shown in \cite{Oliveira20}
this mechanism is trivially extended to any system with a group of
``idle" energy levels \footnote{Actually this condition can be relaxed allowing
for the levels to slightly change during the adiabatic; see suppl. of \cite{Oliveira20}}.

Thus, contrary to what has been suggested for the $XXX$ model \cite{ das2019measurement}, the efficiency increase is not related to any quantum correlation between the spins, but only to the particular structure of the energy levels; which of course depends on the coupling. Note that this is valid
for the two-bath and the measurement engine with the $XXZ$ model
as the working substance. 

We will now illustrate this for some particular cases of the XXZ model.

\subsection{Two-bath engine}

For the two-bath engine the general expression for efficiency is
already too cumbersome to provide any intuition, and it is not shown.

\begin{figure}[h]
\center
\subfigure[]{\includegraphics[scale=0.22]{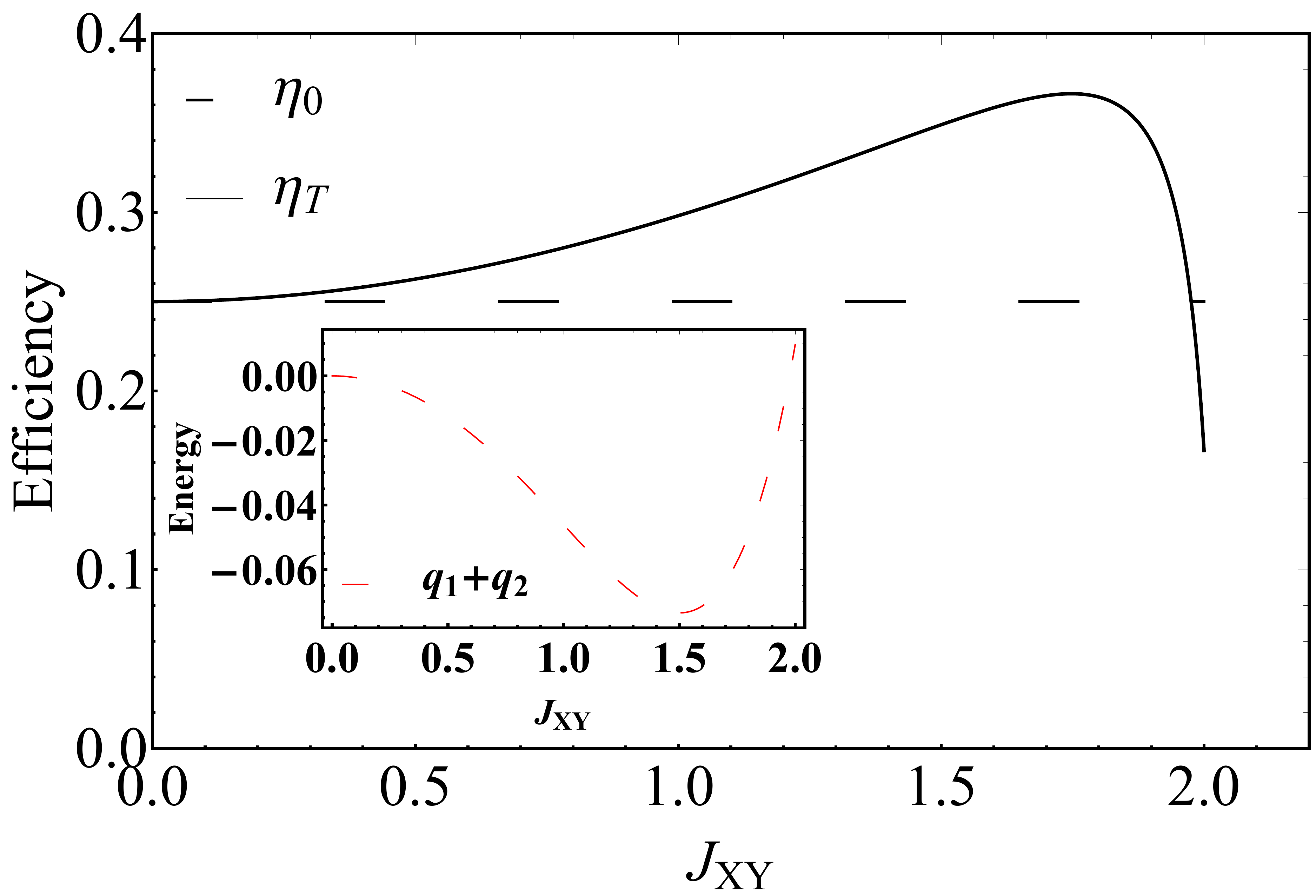}}
\subfigure[]{\includegraphics[scale=0.22]{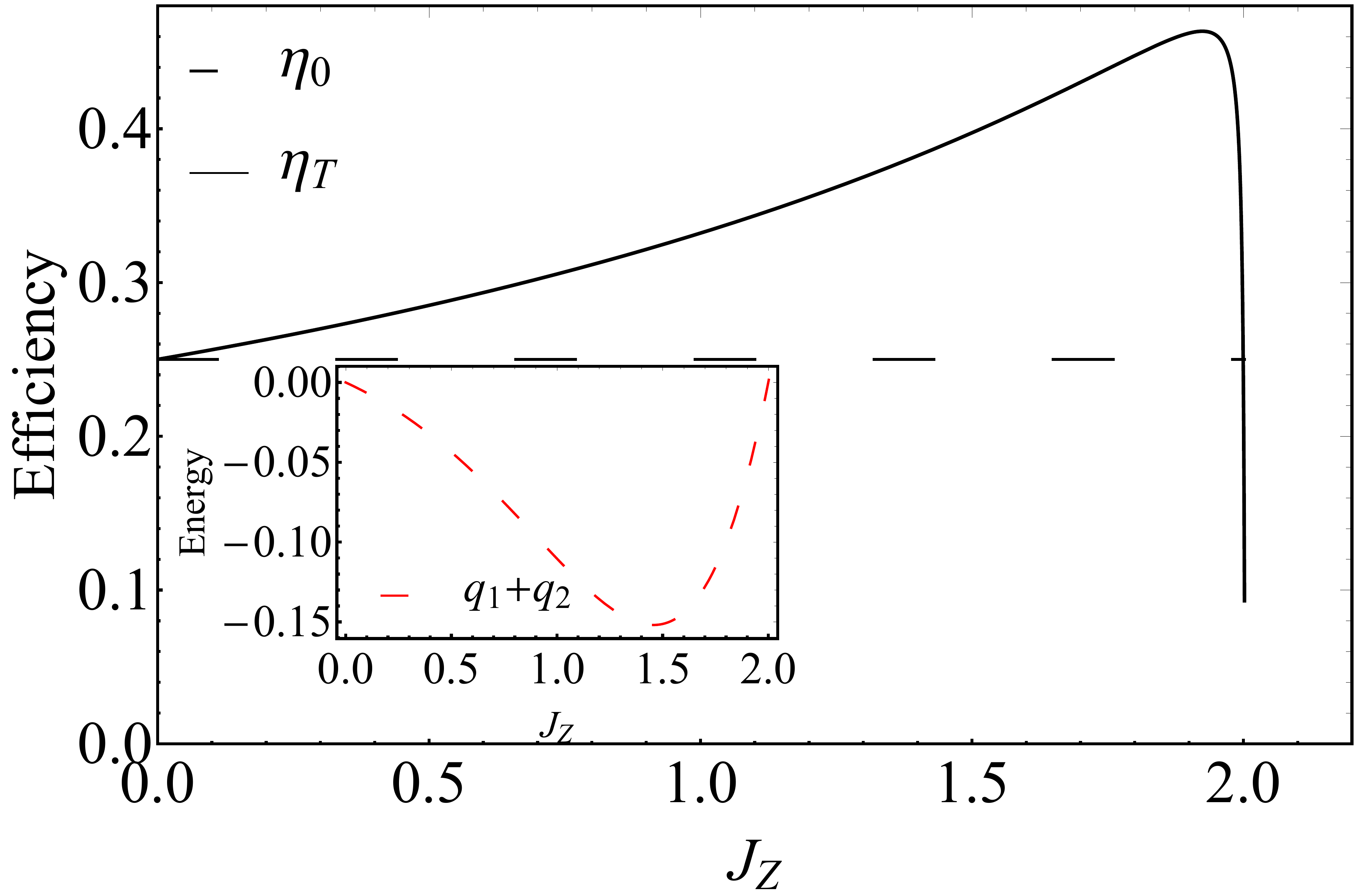}}
\caption{Efficiency of two-bath quantum heat engine for
uncoupled spins ($\eta_0$, dashed black) and two interacting spins ($\eta_T$, solid black). (Inset) We also
plot the heat flowing through the levels that
do not depend on $B$, $q_1+q_2$ (dashed red). In (a)
we have $J_z = 0$ (XX model) and in (b) $J_{xy} = 0$ (transverse field Ising model). In both cases we
see that the increase in the efficiency is due
to  $q_1+q_2<0$.}
\label{hei-baths}
\end{figure}

We first consider the isotropic case with $J_z = 0$, which is the XX model.
In Fig.\ref{hei-baths} (a) we plot the efficiency for the coupled
and uncoupled cases, and the heat flowing through the ``idle" levels.
It can be seen that the increase in the efficiency
is due to $q_1 +q_2 < 0$. This is also illustrated
for the case  $J_{xy} = 0$, the transverse Ising model,
in Fig.\ref{hei-baths} (b). We see the same mechanism
for efficiency increase, which is not related to any quantum
correlations.

\subsection{Measurement-based engine}
\label{sec:Heisenberg-meas}

Now we will consider the measurement engine in the same
particular cases of the two bath engine with two interacting
1/2 spins.  

We use local spin projective measurements on each particle.
In this case, the projectors are given by 
\begin{equation}
\begin{aligned}
M_1 = \ket{+^n} \bra{+^n} \otimes  \ket{+^m} \bra{+^m}, \\
M_2 = \ket{+^n} \bra{+^n} \otimes  \ket{-^m} \bra{-^m}, \\
M_3 = \ket{-^n} \bra{-^n} \otimes  \ket{+^m} \bra{+^m}, \\
M_4 = \ket{-^n} \bra{-^n} \otimes  \ket{-^m} \bra{-^m},
\end{aligned}
\end{equation}
where $\ket{\pm^n} \bra{\pm^n}$ are the projectors for the observable $\vec{\sigma}.\hat{n}$ for one spin and $\ket{\pm^m} \bra{\pm^m}$ are the projectors for the observable $\vec{\sigma}.\hat{m}$ for the other one. With these operators, we can measure each qubit in any direction.

As mentioned, the XXX model was already studied for the measurement
engine \cite{das2019measurement} with precisely these spins measurements.
It was suggested that the
quantum correlation between the spins might be responsible for
the increase in the efficiency in relation to the engine with an uncoupled spin.
We now illustrate, with the same particular examples used in the two-bath models, our results showing that correlations are not the origin of the increase in the efficiency.

\begin{figure}[!ht]
\center
\includegraphics[scale=0.28]{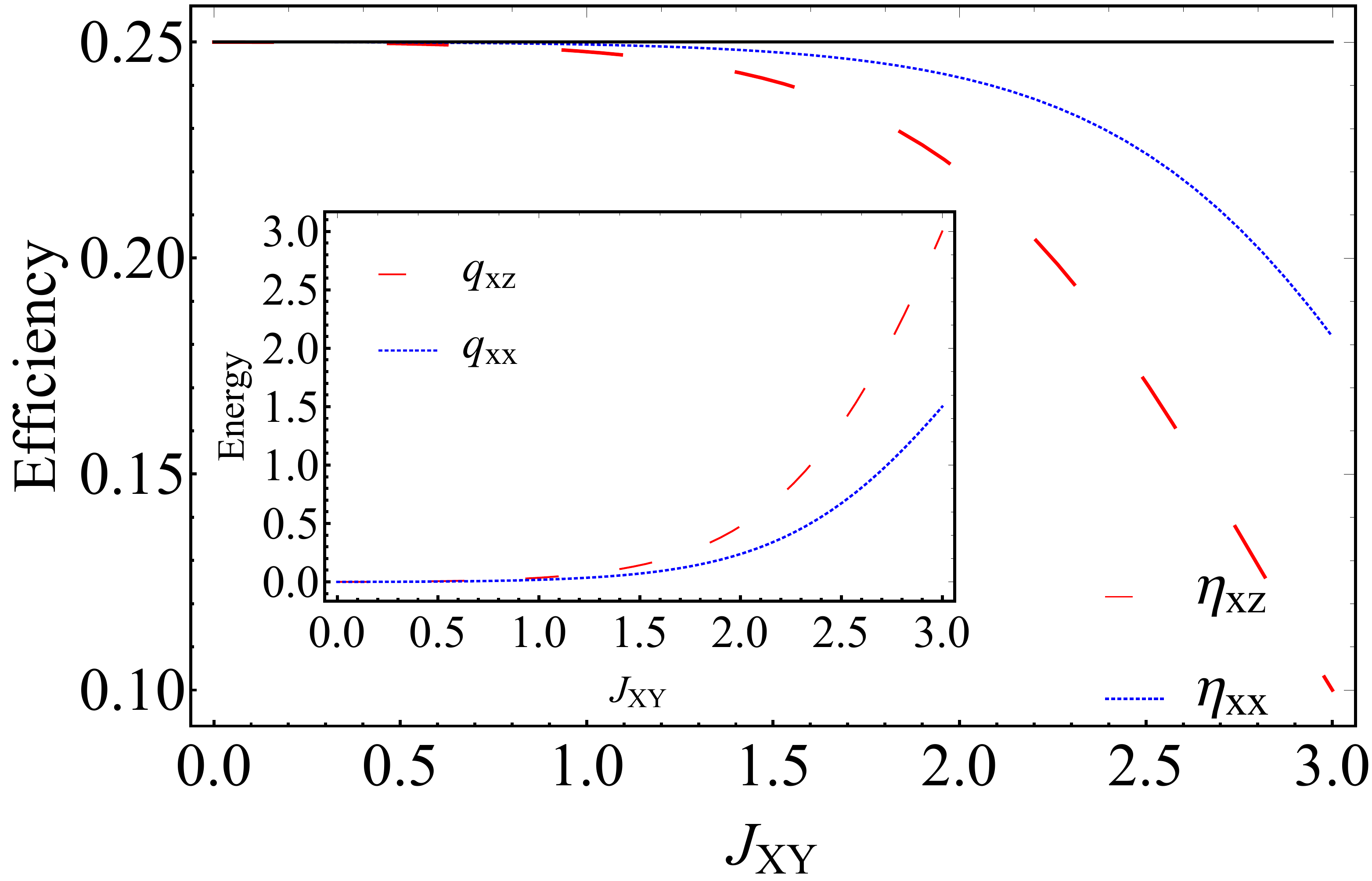}
\caption{Efficiency of measurement engine for two spins 1/2 in the
XX model ($J_z = 0$). The measures are spin measurement in the directions 
$\{\vec{n}=\vec{x}$,$\vec{m}=\vec{z}\}$ (second dashed red) and $\{\vec{n}=\vec{x}$,$\vec{m}=\vec{x}\}$ (first dotted blue).
We also show as a horizontal line the efficiency for a single spin.
In the inset we show $q_1+q_2$ for the xz-direction (dashed red) and xx-direction (dotted blue).}
\label{hei-baths3}
\end{figure}

\begin{figure}[h]
\center
\includegraphics[scale=0.28]{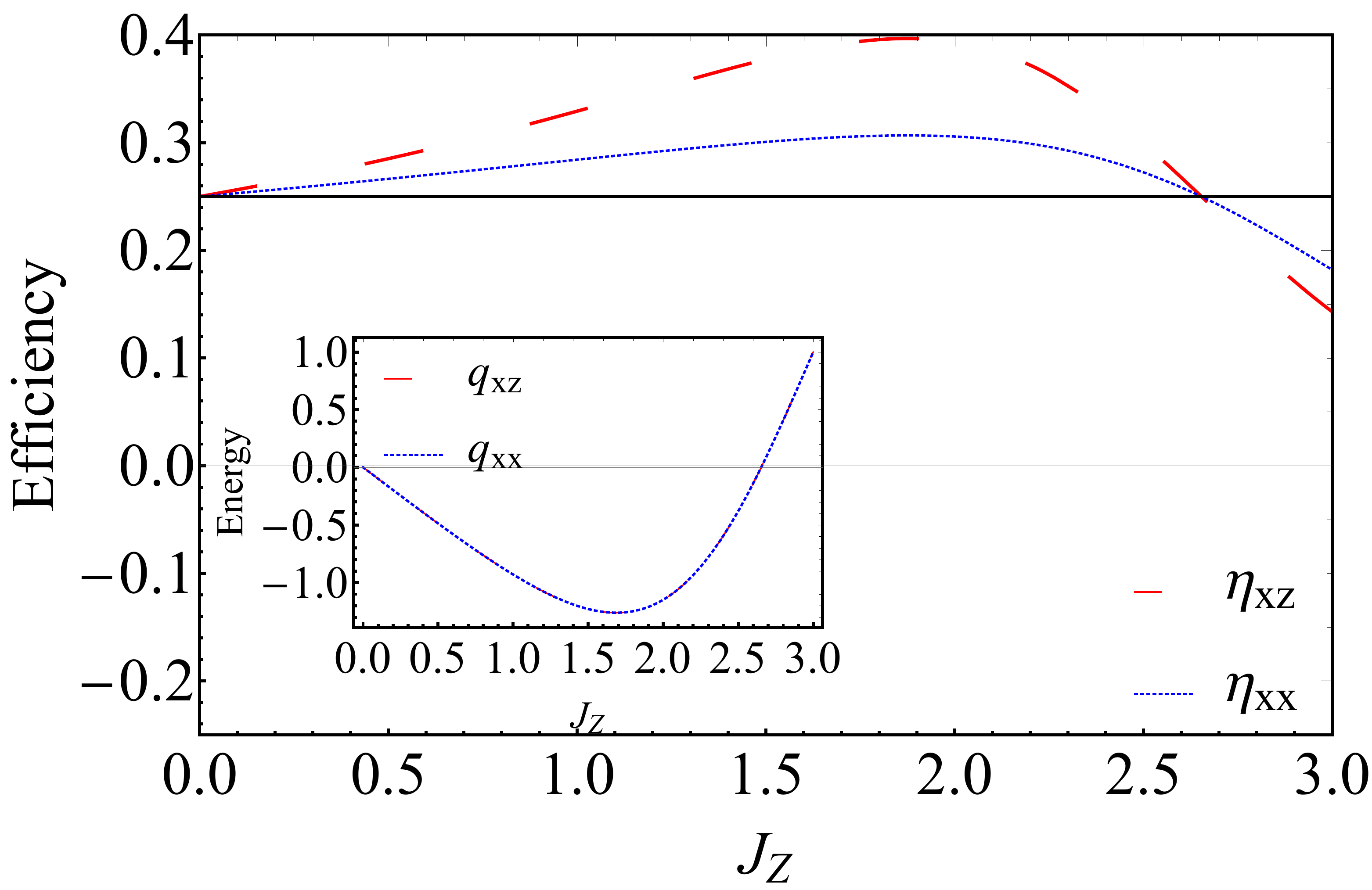}
\caption{Efficiency of measurement engine for two spins 1/2 in the
Ising model ($J_{xy} = 0$). The measures are spin measurement in the directions 
$\{\vec{n}=\vec{x}$,$\vec{m}=\vec{z}\}$ (first curve, dashed red) and $\{\vec{n}=\vec{x}$,$\vec{m}=\vec{x}\}$ (second curve, dotted blue).
Again the horizontal line is the efficiency for uncoupled spins. In the inset we show $q_1+q_2$ for measurements in the xz-direction (dashed red) and the xx-direction (dotted blue)  (the value is equal in both cases)}
\label{hei-baths4}
\end{figure}

As there are many possible measurement directions, no simple
general expression for the efficiency is available. 
We will consider two possible choices: $\{\vec{n}=\vec{x}$,$\vec{m}=\vec{z}\}$ and $\{\vec{n}=\vec{x}$,$\vec{m}=\vec{x}\}$. In Fig. \ref{hei-baths3} we
show the data for the XX model, and in Fig. \ref{hei-baths4} we show the data for the Transverse field Ising Model. We can see for the XX model that the efficiency always decreases for the two chosen measurements, while in the
XXX model it increases \cite{das2019measurement}. We can also see,
in the insets, how the increase in the efficiency only occurs when heat flows from the cold to the hot bath through the ``idle" levels; $q_1+q_2 < 0$.

In summary, we were able to explain the efficiency gain in both the two-bath and the measurement-based models, without invoking entanglement or any other quantum correlation. The efficiency gain only depends on the energy structure of the system,i.e.,
the presence of ``idle" levels.

\section{Conclusion}

In this work we aimed to analyze the mechanism for a performance increase in measurement-based engines with coupled qubits, 
in relation to uncoupled qubits and also to two-bath engines
with coupled qubits. To this end, we studied the Otto
cycle with a three-level system. We showed that in both models, measurement-based and two-baths, the change in the efficiency is due to the structure of the energy levels, more precisely due to one of the levels, $-J$, not changing in the adiabatic process (being an ``idle" level). If one considers the contribution of each energy level to the heats, the increase in the efficiency occurs only when the flow of energy though the $-J$ level is from the cold to the hot bath, something proposed by some
of us in \cite{Oliveira20}. We also showed that this mechanism is
the one responsible for the increase in the efficiency when the engine system
is two spins 1/2 coupled by a Heinsenberg interaction. Thus for the
two types of engines, the efficiency gain is not due to any quantum
correlation, as has been suggested \cite{das2019measurement}.

For the measurement engine, we saw that there is no simple
expression for the efficiency. We thus numerically studied
the efficiency for a general SU(3) projective measurement and
found protocols where it can be higher than that for the two-bath models.
We showed that the efficiency can even approach one, but
with the work output going to zero. The second law is not violated, since the Carnot bound does not apply to engines using a non-thermal energy source, as is the case here.

\begin{acknowledgments}

The authors would like to thank Marcelo França Santos for helpful comments. We acknowledge the Brazilian agencies CNPq and CAPES for financial support. This work is also supported by the Instituto Nacional de Ciência e Tecnologia de Informação Quântica (465469/2014-0). T.R.O. acknowedges the financial support of the Air Force Office of Scientific
600 Research under Award No. FA9550-19-1-0361.

\end{acknowledgments}

\appendix

\section{Proof of Theorem 1} \label{sec:appendA}

\textbf{Theorem 1}: If $\mathcal{E}$ is a unital trace-preserving quantum channel and $\rho$ is a passive quantum state with respect to Hamiltonian $\mathcal{H}$, then 
\begin{align}
\left\langle \Delta E \right\rangle \equiv  \text{Tr} \left[ (\mathcal{E}(\rho)-\rho) \mathcal{H} \right]\geq0.\label{eq:deltaE1}
\end{align}
\smallskip

\emph{Proof:} Let $\{\ket{n}\}$ be the eigenvectors of $\mathcal{H}$, with corresponding energies $E_{n}$, numbered in nondecreasing order. Also, let $\{M_{\alpha}\}$, satisfying $\sum_{\alpha} M_{\alpha} M_{\alpha}^{\dagger} = \sum_{\alpha}  M_{\alpha}^{\dagger}M_{\alpha} = \mathds{1}$ be a set of Kraus operators for $\mathcal{E}$.

Define a square matrix $T$ with elements
\begin{align}
 T_{mn} = \sum_{\alpha} \left|\bra{m} M_{\alpha}\ket{n}\right|^{2}
\end{align}
Note that, for any trace-preserving quantum channel,
\begin{align}
 \sum_{m} T_{mn} &= \sum_{\alpha} \bra{n}  M_{\alpha}^{\dagger} \sum_{m}\ketbra{m}{m} M_{\alpha} \ket{n} = 1 
\end{align}
so $T$ is a \emph{stochastic} (or `Markov') matrix. $T$ can be interpreted as the transfer matrix mapping the original probability distribution for energy, $p_{n} = \bra{n}\rho\ket{n}$ to the new one after the measurement:
\begin{align}
 p'_{m} & \equiv \bra{m}\mathcal{E}(\rho)\ket{m}= \sum_{\alpha} \bra{m} M_{\alpha} \sum_{n} p_{n} \ketbra{n} {n}  M_{\alpha}^{\dagger} \ket{m} \nonumber \\
  &= \sum_{n} T_{mn} p_{n}.
\end{align}
In other words, $T_{mn}$ is the conditional probability $p(m|n)$ of having energy $E_{m}$ after the measurement is performed, given that we had energy $E_{n}$ before. 

It is convenient at this point to define probability and energy vectors $\vec{p} = (p_{1}, p_{2}, \ldots,)$ and  $\vec{E}=(E_{1}, E_{2},  \ldots, )$, 
where $\vec{E}$  is in ascending order. The change in average energy, $\left\langle \Delta E \right\rangle$, can then be written
\begin{align}
\left\langle \Delta E \right\rangle  = \vec{E} \cdot (T - \mathds{1}) \vec{p} \label{eq:deltaE2}
\end{align}

In Ref. \cite{yi2017single}, $\mathcal{E}$ was restricted to the class of measurements where $M_{\alpha}$ can all be chosen to be Hermitian. It was shown that in that case $T$ is a symmetric matrix, a fact that was then exploited to prove Eq.\,(\ref{eq:deltaE1}). 
Here we have imposed the weaker condition that $\mathcal{E}$ is \emph{unital}. Nevertheless, Eq.\,(\ref{eq:deltaE1}) still holds. To see this, note that
\begin{align}
 \sum_{n} T_{mn} &= \sum_{\alpha} \bra{m} M_{\alpha} \sum_{n}\ketbra{n}{n}  M_{\alpha}^{\dagger} \ket{m} \nonumber \\ &=  \bra{m}\left( \sum_{\alpha} M_{\alpha}M_{\alpha}^{\dagger}  \right) \ket{m} = 1.
\end{align}
In other words, for unital $\mathcal{E}$, $T$ is in fact a \emph{bistochastic} matrix. These matrices have many special properties linked to the concept of majorization \cite{olkin1979inequalities, bhatia1997matrix}. In particular: \emph{Birkhoff's Theorem} states that a square matrix $T$ is bistochastic if and only if it can be written as a convex combination of permutation matrices: $T = \sum_{j} q_{j} \sigma_{j}$, where $\sigma_{j}$ are permutations, $\sum_{j} q_{j} = 1$ and $0 < q_{j} \leq 1$. In order to prove that Eq.\,(\ref{eq:deltaE2}) is $\geq0$, it suffices therefore to show that, for any permutation matrix $\sigma$, $\vec{E} \cdot (\sigma \vec{p})  \geq  \vec{E} \cdot \vec{p} $. 

This follows from the fact that $\rho$ is `passive', which means that $\vec{p}$ is ordered in nonincreasing order. Energies and probabilities are therefore perfectly anticorrelated, with the greatest probabilities matching the smallest energies. It is then intuitively clear that any rearrangement of the probability vector $\vec{p}$ will increase the average energy. This statement can be made precise using the mathematical result known as the `rearrangement inequality' (see, e.g., \cite{hardy1952inequalities},  Section 10.2, Theorem 368). \medskip

\section{Qutrit efficiency}

Here we give expressions for the efficiency for the measurement-based engine in the Heisenberg model, for the three chosen set of angles studied in section \ref{sec:Heisenberg-meas}

\begin{equation}
\eta_1 = \dfrac{(B_f - B_i) +  \dfrac{(9.1e^{J+B_i})10^{-16}}{(0.19 -1.17e^{2B_i} + 0.97e^{J+B_i})} J}{B_f + \dfrac{(0.025 + e^{2B_i} -1.02e^{J+B_i})}{(0.19 -1.17e^{2B_i} + 0.97e^{J+B_i})}J},
\end{equation}

\begin{equation}
\eta_2 = \dfrac{(B_f - B_i) +  \dfrac{(e^{J+B_i} -0.12e^{2B_i}-0.01)10^{-15}}{(0.92 -1.79e^{2B_i} + 0.87e^{J+B_i})} J}{B_f + \dfrac{(0.13 + e^{2B_i} -1.13e^{J+B_i})}{(0.92 -1.79e^{2B_i} + 0.87e^{J+B_i})}J},
\end{equation}

\begin{equation}
\eta_3 = \dfrac{(B_f - B_i) - \dfrac{(4.54 + 0.14e^{2B_i} -9.07e^{J+B_i})10^{-15}}{(46.54 -7.68e^{2B_i} - 38.86e^{J+B_i})} J}{B_f + \dfrac{(39.86 + e^{2B_i} - 40.86e^{J+B_i})}{(46.54 -7.68e^{2B_i} - 38.86e^{J+B_i})}J},
\end{equation}
where $\eta_1$ is the efficiency with the angles given by $\theta=\phi=0.7\pi$ and $\psi=\chi=0.5\pi$, $\eta_2$ with $\theta=\phi=\chi= 0.7\pi$ and $\psi=0.5\pi$; and $\eta_3$ with $\theta=\phi=\chi=\psi=0.3\pi$.

\bibliography{Measurement-based_quantum_heat_engine_in_a_multilevel_system}

\end{document}